\documentclass[iop]{emulateapj}

\newcommand{\chandra} {{\it Chandra}}
\newcommand{\nustar} {{\it NuSTAR}}
\newcommand{\swiftxrt} {{\it Swift}/XRT}

\newcommand{\fx} {$F_{\rm{X}}$}
\newcommand{\chisq} {$\chi^2$}

\newcommand{\degree}{{$^\circ$}}
\newcommand{\ergs}{\mbox{\thinspace erg\thinspace s$^{-1}$}}
\newcommand{\ergcms}{\mbox{\thinspace erg\thinspace cm$^{-2}$\thinspace s$^{-1}$}}

\newcommand{\msol} {$M_{\odot}$}

\shorttitle{A 63.8-day periodic flux modulation from the ULX pulsar in M82.}
\shortauthors{Brightman et al.}

\begin{document}

\title{A $\sim60$-day super-orbital period originating from the ultraluminous X-ray pulsar in M82}

\author{Murray Brightman$^{1}$, Fiona A. Harrison$^{1}$, Matteo Bachetti$^{2,1}$, Yanjun Xu$^{1}$, Felix F\"{u}rst$^{3}$, Dominic J. Walton$^{4}$, Andrew Ptak$^{5,6}$, Mihoko Yukita$^{5,6}$, Andreas Zezas$^{7,8,9}$}

\affil{$^{1}$Cahill Center for Astrophysics, California Institute of Technology, 1216 East California Boulevard, Pasadena, CA 91125, USA\\
$^{2}$INAF, Osservatorio Astronomico di Cagliari, Via della Scienza 5, 09047 Selargius, Italy\\
$^{3}$European Space Astronomy Centre (ESA/ESAC), Operations Department, Villanueva de la Ca\~{n}ada (Madrid), Spain\\
$^{4}$Institute of Astronomy, Madingley Road, Cambridge CB3 0HA, UK\\
$^{5}$Johns Hopkins University, Homewood Campus, Baltimore, MD 21218, USA\\
$^{6}$NASA Goddard Space Flight Center, Greenbelt, MD 20771, USA\\
$^{7}$Department of Physics, and Institute for Theoretical and Computational Physics, University of Crete, Heraklion 71003, Greece\\
$^{8}$Foundation for Research and Technology ? Hellas (FORTH), Heraklion 71003, Greece\\
$^{9}$Harvard-Smithsonian Center for Astrophysics, 60 Garden St., Cambridge, MA 02138, USA\\}

\begin{abstract}

Ultraluminous X-ray (ULX) pulsars are a new class of object powered by apparent super-critical accretion onto magnetized neutron stars. Three sources in this class identified so far; M82~X-2, NGC~5907~ULX-1 and NGC~7793~P13, have been found to have two properties in common; $\sim1$-s spin periods, and for NGC~5907~ULX-1 and NGC~7793~P13 periodic X-ray flux modulations on timescales of $\sim60-80$ days. M82 X-2 resides in a crowded field that includes the ULX M82 X-1 separated from X-2 by 5\arcsec, as well as other bright point sources. A 60-day modulation has been observed from the region but the origin has been difficult to identify; both M82 X-1 and X-2 have been suggested as the source. In this paper we present the analysis of a systematic monitoring campaign by \chandra, the only X-ray telescope capable of resolving the crowded field.  From a simple Lomb-Scargle periodogram analysis and a more sophisticated Gaussian Process analysis we find that only X-2 exhibits a periodic signal around 60 days supporting previous claims that it is the origin. We also construct a phase-averaged flux profile of the modulations from higher cadence \swiftxrt\ data and find that the flux variations in the \chandra\ data are fully consistent with the flux profile. Since the orbit of the neutron star and its companion is known to be 2.5 days, the $\sim60$-day period must be super-orbital in origin. The flux of the modulations varies by a factor of $\sim$100 from minimum to maximum, with no evidence for spectral variations, making the origin difficult to explain.

\end{abstract}


\section{Introduction}

The ultraluminous X-ray source (ULX) M82 X-2 is the second brightest point source in the galaxy M82 \citep{matsumoto01,feng07,kong07} and was recently discovered to be powered by accretion on to a magnetized neutron star \citep[][B14]{bachetti14} from the detection of coherent X-ray pulsations by \nustar\ \citep{harrison13} while observing the supernova 2014J. This discovery revolutionized the field of ULXs, which for long time were thought to be powered by black holes because of their large apparent luminosities. 

The pulsar has a spin period of 1.37\,s and the pulse profile is close to sinusoidal (B14). The orbit of the neutron star around its companion star, which has a minimum mass of 5.2 \msol, is close to circular, has a period of 2.5 days and a projected semimajor axis of $\sim7$ million km. A linear spin up is also observed from the pulsations during the \nustar\ observations, with a pulse derivative $\dot{P}\simeq-2\times10^{-10}$ s s$^{-1}$, that varies from observation to observation.

Since X-2 is separated from its brighter neighbor X-1 \citep{kaaret06} by only 5\arcsec\ on the sky, \chandra\ is the only X-ray telescope capable of resolving the two sources. In \cite{brightman16} we analyzed the archival \chandra\ observations of X-2 finding that the luminosity of the source is observed to range from $\sim10^{38}$ \ergs\ $\sim10^{40}$ \ergs. Its spectrum can be described in the \chandra\ band by an absorbed power-law with $\Gamma=1.33\pm0.15$, typical of other known pulsars. From \nustar\ data we isolated the pulsed emission which was best fit by a power-law with a high-energy cut-off, where $\Gamma=0.6\pm0.3$ and $E_{\rm C}=14^{+5}_{-3}$ keV.

Since the discovery that the ULX M82 X-2 is powered by an accreting neutron star, intense theoretical work to determine how a pulsar can sustain such extreme luminosities has followed \citep[e.g.][]{eksi15, dallosso15, mushtukov15, tsygankov16, dallosso16, karino16,kawashima16,king16}. In addition, three further ULX pulsars have recently been identified, one in NGC~5907 \citep[ULX-1,][]{israel17a} a second in NGC~7793 \citep[P13,][]{fuerst16,israel17b} and a third in NGC~300 \citep[ULX1/SN2010da,][]{carpano18}. Furthermore, another ULX was identified as being powered by a neutron star from the detection of a likely cyclotron resonance scattering feature, albeit with no pulsations detected, in M51 \citep[ULX8,][]{brightman18}. These all reside in more isolated environments than M82 X-2 making them easier to study. 

The pulsars have remarkably similar basic properties: peak luminosities that exceed their Eddington limits by a factor of 50--500, spin periods of order 1\,s and  periodic variability in their fluxes on timescales of tens-of-days. NGC~5907~ULX-1 was found to have a periodic modulation in its X-ray flux of 78 days \citep{walton16b} and NGC~7793~P13 has a $\sim64-65$-day periodicity in the optical \citep{motch14} and X-rays \citep{hu17}. While several claims have been made that M82 X-2 exhibits a $\sim60$-day periodic flux modulation \citep{qiu15,kong16}, due to the crowded field of M82, the periodicity from X-2 is less certain.

The periodic X-ray variability of the central region of M82 has been well studied, originating from the discovery of a 62-day periodic modulation in {\it RXTE} data by \cite{kaaret06} which was at first interpreted as being due to M82 X-1 and orbital in nature. Following this, the periodic signal was observed to change phase \citep{pasham13}, which the authors considered to be more likely due to a super-orbital origin. Later analysis of \swiftxrt\ data by \cite{qiu15} implied that the periodicity does not in fact come from M82 X-1, rather one of three point sources $\sim5$\arcsec\ to the south east, one of which is the ULX pulsar, X-2. They also confirm that the period is not stable, finding that it changes in phase. Most recently, \cite{kong16} analyzed all archival \chandra\ data identifying a possible 55-day period originating from X-2. In their own analysis of the \swiftxrt\ data, they find that the period varies between 55--62 days. All of these authors concluded that a systematic monitoring campaign by \chandra\ was necessary to correctly identify the origin of the periodicity.

\begin{figure}
\begin{center}
\includegraphics[width=90mm]{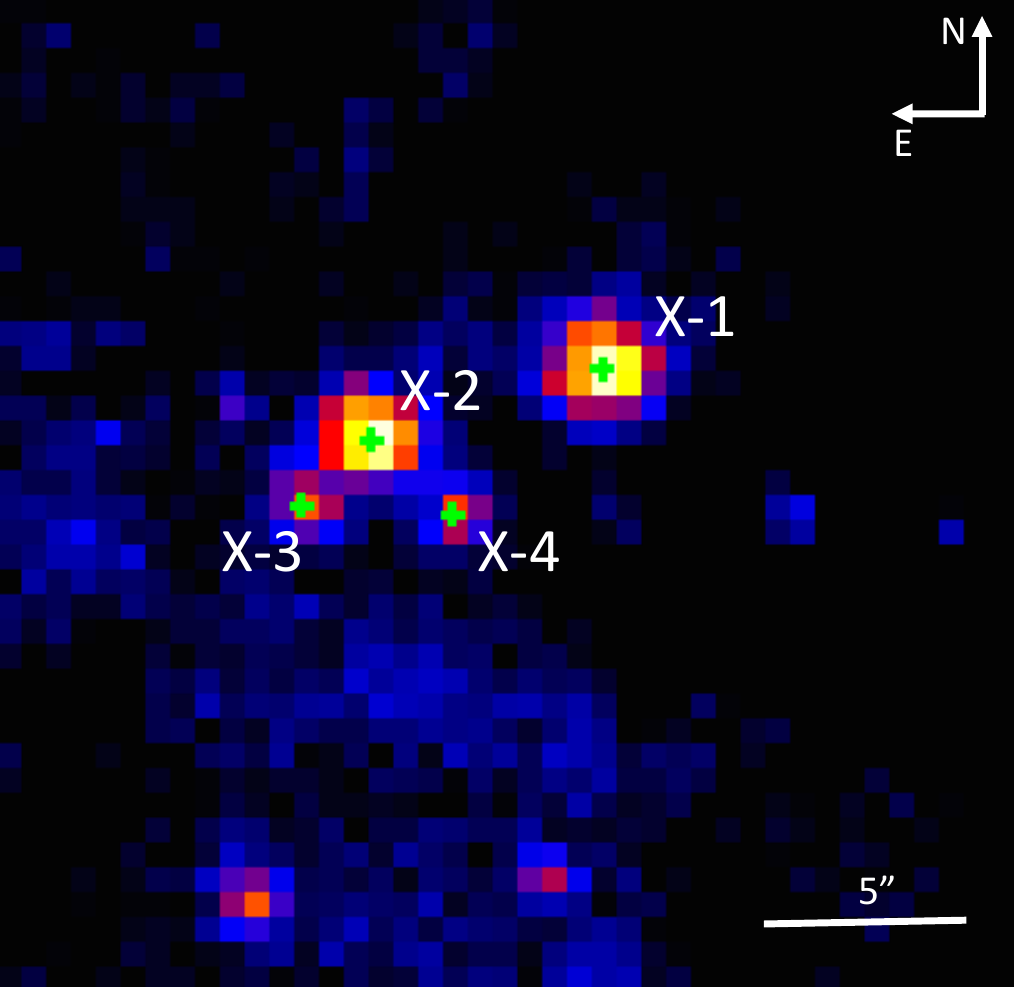}
\caption{On-axis \chandra\ image of the ULXs in M82 (obsID 16580) with the four sources of interest marked with green `+'  and labeled.}
\label{fig_16580_img}
\end{center}
\end{figure}

In this paper we present analysis of a new systematic monitoring campaign on M82 by \chandra\ which took place in 2016. The primary goals of this campaign were to perform a temporal analysis of X-1 and X-2 to search for orbital and super-orbital modulations; to perform spectroscopic studies of the ULXs and to study the nature of the other binary systems in M82. We focus here on unambiguously determining the source of the $\sim60$-day periodic signal. We concentrate on the four bright X-ray sources that have been claimed to be the possible source of the signal, X-1, X-2, X-3 (CXOU J095551.2+694044) and X-4 (CXOU J095550.6+694944), shown in Figure \ref{fig_16580_img}. We begin in Section \ref{sec_swift} where we conduct our own analysis of the \swiftxrt\ data presented by \cite{qiu15} and \cite{kong16} to use as a baseline for our \chandra\ study. In Section \ref{sec_chandra} we present the analysis of our new \chandra\ data and in Section \ref{sec_period} we combine these results to determine the source of the periodicity. In section \ref{sec_implications} we discuss the implications of our results and summarize and conclude in Section \ref{sec_conclusions}.

\section{Swift/XRT data analysis}
\label{sec_swift}

While \swiftxrt, which has a point spread function (PSF) of 18\arcsec\ \citep[half-power diameter at 1.5 keV,][]{moretti05}, cannot resolve the various point sources in the center M82, it has conducted monitoring of the galaxy with a typical cadence of a few days between 2012--2016. Since this monitoring ran contemporaneously with our \chandra\ observations, and for four years prior, it provides us with a valuable baseline for our \chandra\ study. A total of 227 observations have been made of the galaxy over the 5-year period which we use to calculate a long-term lightcurve. 

We calculate the fluxes via spectral fitting. We use the {\sc heasoft} (v 6.16) tool {\sc xselect} to filter events from a 49\arcsec\ radius region centered on the ULXs and to extract the spectrum. This extraction region encloses all sources of X-ray emission in the galaxy. Background events were extracted from a nearby circular region of the same size. We group the spectra with a minimum of one count per bin using the {\sc heasoft} tool {\sc grppha}. We conduct spectral fitting using {\sc xspec} v12.8.2 in the range 0.2--10 keV. We fit the spectra with a simple power-law subjected to absorption intrinsic to M82 at z=0.00067 ({\tt zwabs*powerlaw} in {\sc xspec}) with the Cash statistic \citep{cash79} which uses a Poisson likelihood function and is hence most suitable for low numbers of counts per bin. From this model we calculate the observed flux in the 0.5--8 keV range, equivalent to the \chandra\ band. 

\begin{figure*}
\begin{center}
\includegraphics[width=180mm]{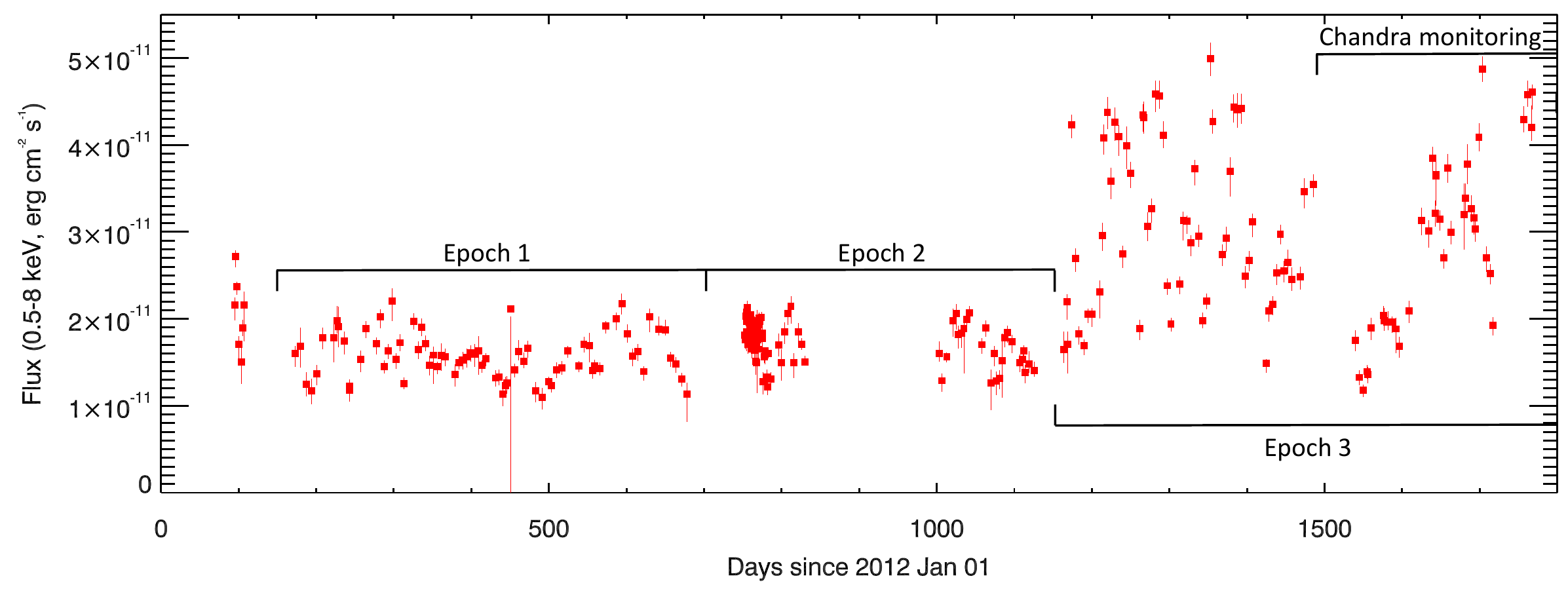}
\caption{\swiftxrt\ 0.5--8 keV observed flux of M82 during the period 2012--2016. We split these data into three epochs, labeled, for timing analysis. We also mark the period over which the \chandra\ data were taken which was contemporaneous with the \swiftxrt\ monitoring.}
\label{fig_xrt_ltcrv}
\end{center}
\end{figure*}

Figure \ref{fig_xrt_ltcrv} shows the flux as a function of the number of days since 2012-01-01. While the X-ray emission remained at 1--2$\times10^{-11}$ \ergcms\ for much of the period covered, the activity increased up to $\sim5\times10^{-11}$ \ergcms\ after $\sim1150$ days and for the rest of the period due to a flaring episode from X-1 \citep{brightman16c}. 

In order to investigate the periodic signal and any possible variations in phase or period, we split the data into three epochs of approximately the same size. Epoch 1 ranges from 150--700 days where the \swiftxrt\ monitoring was homogeneous and the flux was low. Epoch 2 ranges from 700--1150 days where the \swiftxrt\ data included a period of intense monitoring of the supernova 2014J \citep{fossey14} and a gap of $\sim200$ days and again the flux was low. In their analysis of the same \swiftxrt\ data, \cite{qiu15} excluded the SN 2014J data to avoid contamination from the supernova. However, as \cite{kong16} pointed out, the supernova was not detected in the X-rays below 10 keV \citep[$F_{\rm X}<2.6\times10^{-15}$~\ergcms,][]{margutti14}, so we use all of these data in our analysis. Epoch 3 ranges 1150--1800 days where again the monitoring was homogeneous but the flux was high.

\begin{figure*}
\begin{center}
\includegraphics[width=180mm]{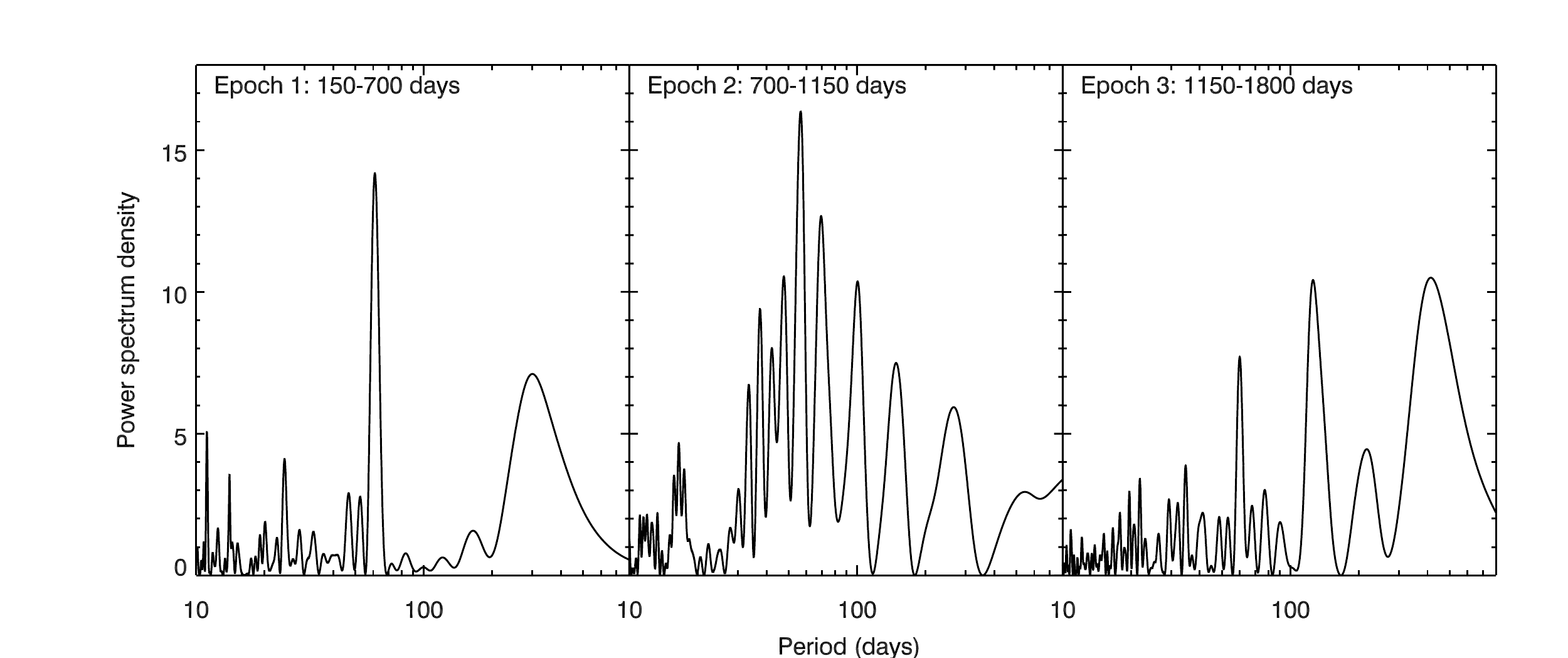}
\caption{Lomb-Scargle periodograms of the \swiftxrt\ data during the periods 150--700 days (left), 700--1150 days (middle) and 1150--1800 (right). Periodic signals that peak at 61.0 days and 56.5 days are found for the first two epochs, but no signal is detected during the final one, during which X-1 is flaring and our \chandra\ monitoring took place.}
\label{fig_xrt_period}
\end{center}
\end{figure*}

We conducted Lomb-Scargle \citep[LS,][]{lomb76,scargle82} periodogram analysis on these three epochs. We search over periods from 10--1000 days with $10^4$ independent frequencies. The resulting periodograms are presented in Figure \ref{fig_xrt_period}. Periodic signals that peak at 61.0 days and 56.5 days are  found in epoch 1 and epoch 2 respectively, but no such strong peak is found from epoch 3. The 61.0-day and 56.5-day periods are consistent with the 62.0$\pm$3.3 and 54.6$\pm$2.1 day periods detected by \cite{qiu15}, determined to each be coming from different sources, either X-2, X-3 or X-4. However, since we detect these different periodicities in different epochs, we assume they come from the same source \citep[e.g.][]{kong16}. \cite{qiu15} also do not detect any significant periodicity from X-1, which agrees with our non-detection of periodicity during the flaring episode of epoch 3. 

\begin{figure}
\begin{center}
\includegraphics[width=90mm]{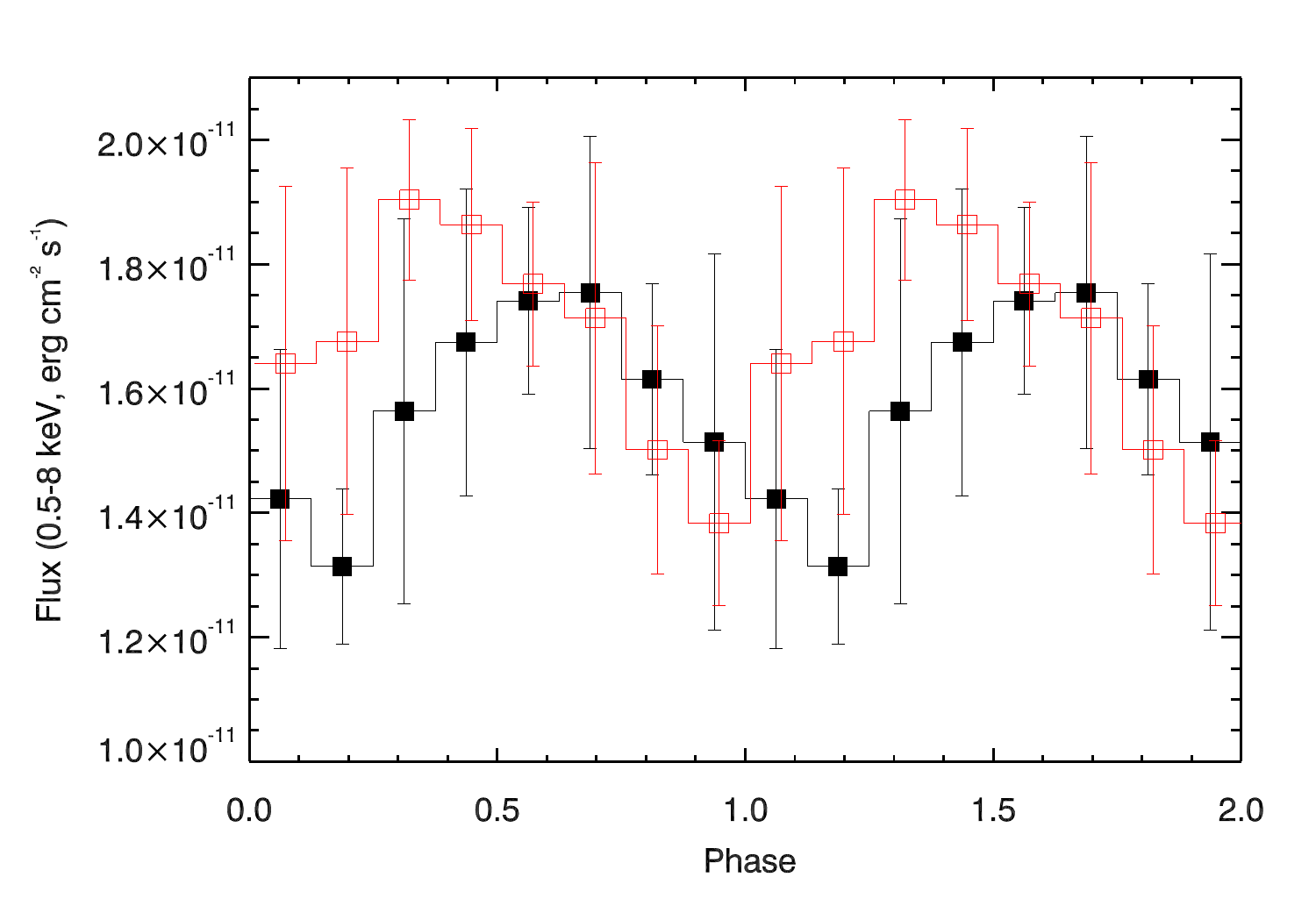}
\caption{The mean flux profiles of the 61.0-day (black filled squares) and 56.5-day (red empty quares, shifted by +0.01 in phase for clarity) signals detected in \swiftxrt\ monitoring during epoch 1 and epoch 2 respectively, with 1$\sigma$ error bars calculated from epoch folding. $T_0$ corresponds to 2012-01-01. Two cycles are shown for clarity.}
\label{fig_xrt_profile}
\end{center}
\end{figure}

We carry out epoch folding of the \swiftxrt\ data in the two epochs over their respective periods to determine the average flux profile of the signals. We assign each \swiftxrt\ data point a phase and average the data over 8 phase bins. The resulting profiles of the modulations are presented in Figure \ref{fig_xrt_profile}, where $T_0$ corresponds to 2012-01-01. The error bars represent the 1$\sigma$ spread in the \swiftxrt\ data. Since the data from the two epochs have been folded on different periods but with the same $T_0$, it appears as if a phase shift has occurred, but this is not necessarily the case.

\section{Chandra data analysis}
\label{sec_chandra}

The majority of the \chandra\ data analyzed here were all taken during 2016 (Cycle 17) as part of a systematic monitoring program on M82. The program consisted of 12 individual observations taken at $\sim$monthly intervals.  We additionally use three archival observations taken in the two years preceding the monitoring campaign to increase the baseline of our analysis. Full details are listed in Table \ref{table_obsdat}. All 2016 observations were taken with ACIS-I at the optical axis with only a 1/8th sub-array of pixels on chip I1 or I3 turned on, depending on the roll angle. The ULXs at the center of M82 were placed 3\arcmin.5 off-axis to smear out the PSF enough to mitigate the effects of pile-up, but not so much as to cause significant blending of the PSFs. The sub-array of pixels was used to decrease the readout time of the detector to 0.4\,s, further mitigating the effects of pile-up.

\begin{table}
\centering
\caption{Chandra observational data}
\label{table_obsdat}
\begin{center}
\begin{tabular}{r c c c c l}
\hline
ObsID	& Date	& Start time	& Exposure \\
(1) & (2) & (3) & (4)  \\
\hline
17578&2015-01-16&13:40:00&10.0\\
16023&2015-01-20&00:26:16&10.0\\
17678&2015-06-21&02:45:09&10.0\\
18062&2016-01-26&19:44:49&      25.1\\
18063&2016-02-24&00:37:06&      25.1\\
18064&2016-04-05&16:04:41&      25.1\\
18068&2016-04-24&20:02:13&      25.1\\
18069&2016-06-03&22:10:58&      25.1\\
18067&2016-07-01&23:18:08&      26.1\\
18065&2016-07-29&07:50:01&      25.1\\
18073&2016-08-19&08:49:59&      40.1\\
18066&2016-09-03&10:11:06&      25.1\\
18070&2016-10-08&00:39:30&      25.1\\
18071&2016-11-03&07:26:45&      25.1\\
18072&2016-12-01&09:35:57&      25.6\\

\hline
\end{tabular}
\tablecomments{Details of the 12 \chandra\ observations of M82 taken in 2016 used in our analysis, plus 3 older observations, ordered by date. Column (1) gives the obsID, column (2) gives the date of the observation, column (3) gives the start time of the observation and column (4) gives the total exposure time in ks.}
\end{center}
\end{table}

\begin{figure}
\begin{center}
\includegraphics[width=90mm]{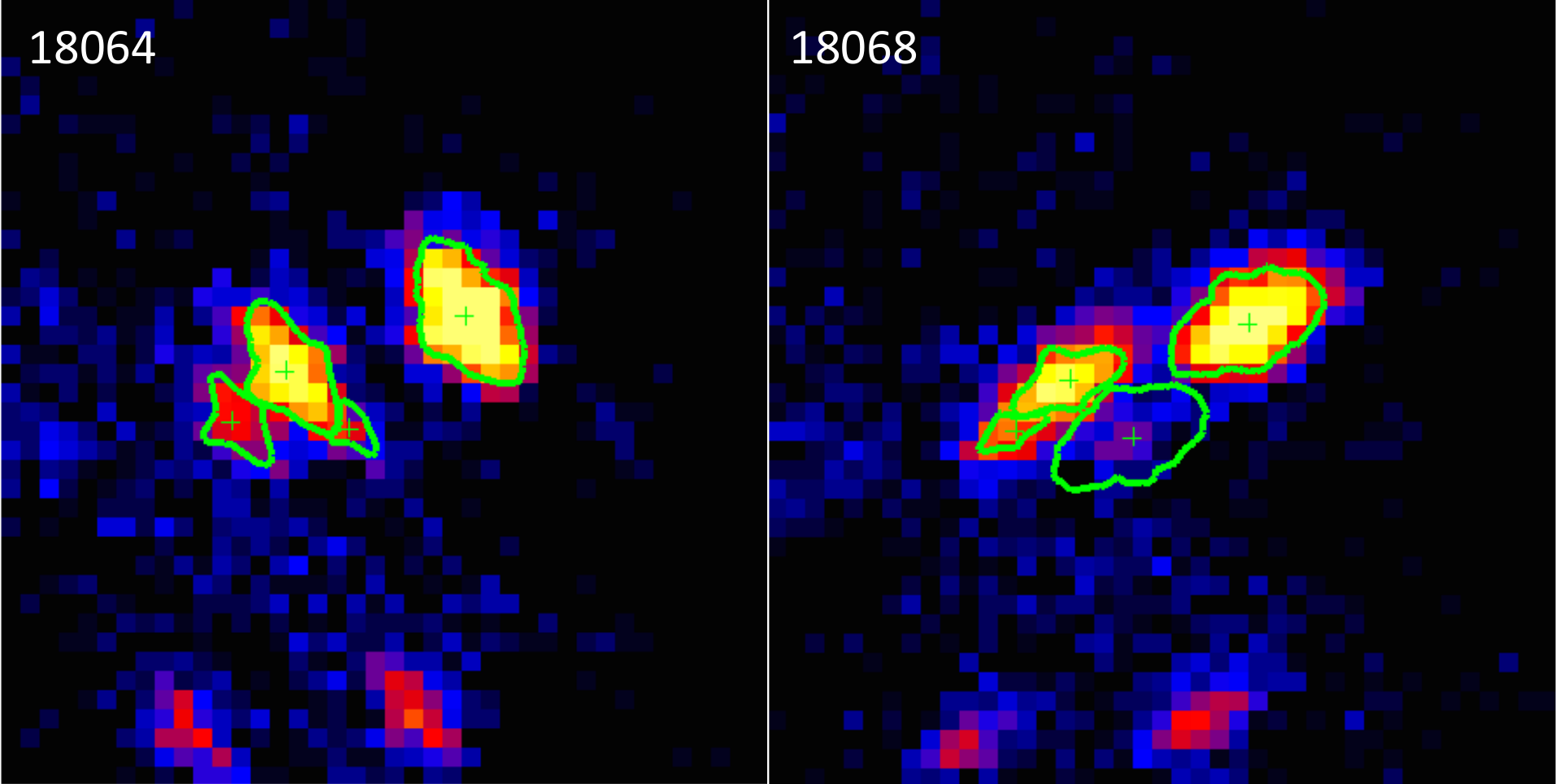}
\caption{Example \chandra\ images of the ULXs in M82 from our observing campaign showing the extraction regions used (obsIDs 18064 and 18068). These have been determined by the {\sc acis\_extract} software which calculates the shape of a region enclosing 90\% of the local PSF at the source position. These are then rescaled depending on their proximity to nearby sources. The scale and orientation are the same as Figure \ref{fig_16580_img}.}
\label{fig_18064+18068_img}
\end{center}
\end{figure}

Due to the off-axis smearing of the PSF and the varying roll angles we use the {\sc acis\_extract} \citep[AE,][]{broos10} software to determine the spectral extraction regions. AE extracts spectral information for each source from each individual observation based on the shape of the local PSF which varies significantly as a function of position on the detector.  We use the known source positions, astrometrically correcting the images by eye (typical shifts of $<1$\arcsec). AE also resizes extraction regions in the case of crowded source positions such that they do not overlap, as is the case for M82. Where possible we use regions where 90\% of the PSF has been enclosed at 1.5 keV. Figure \ref{fig_18064+18068_img} shows an example of two observations with differing roll angles and the extraction regions used that account for the shapes of the PSF at their respective positions. Background spectra are extracted from an events list which has been masked of all the point sources using regions and contain at least 100 counts. We use AE version 2014-08-29, which calls on {\sc ciao} version 4.7 and {\sc CALDB} version 4.6.5.

\begin{figure*}
\begin{center}
\includegraphics[width=180mm]{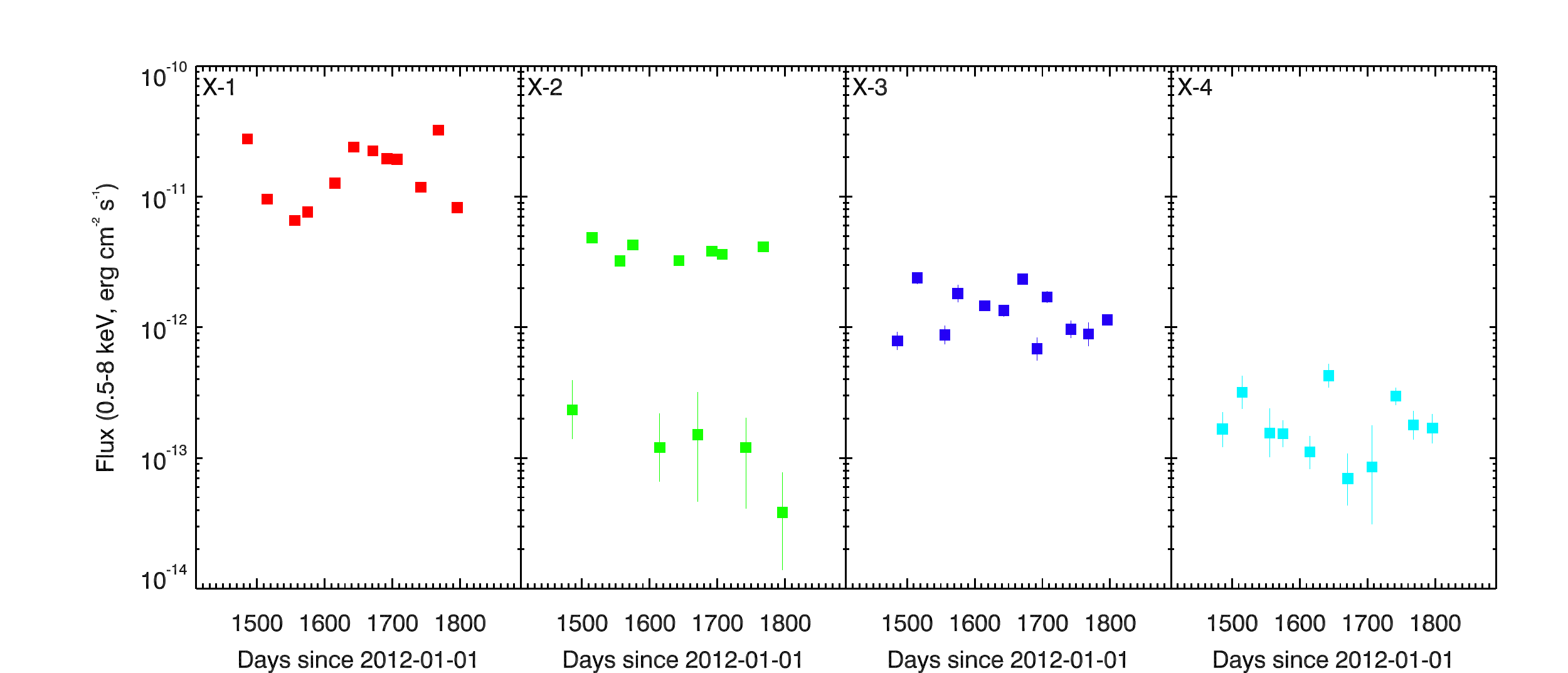}
\caption{\chandra\ 0.5--8 keV observed fluxes of the four bright sources in the center of M82 during 2016. The error bars show the uncertainty in the flux at the 90\% level, which may be smaller than the data points.}
\label{fig_chandra_ltcrv}
\end{center}
\end{figure*}

We follow the same spectral fitting method as for the \swiftxrt\ data, grouping the spectra with a minimum of one count per bin, but in the range 0.5--8 keV appropriate for \chandra. We fit the spectra with a simple absorbed power-law and use a {\tt cflux} component to calculate the observed 0.5--8 keV flux of each source. We plot these fluxes in Figure \ref{fig_chandra_ltcrv} with their 90\% uncertainties.

\begin{figure*}
\begin{center}
\includegraphics[width=180mm]{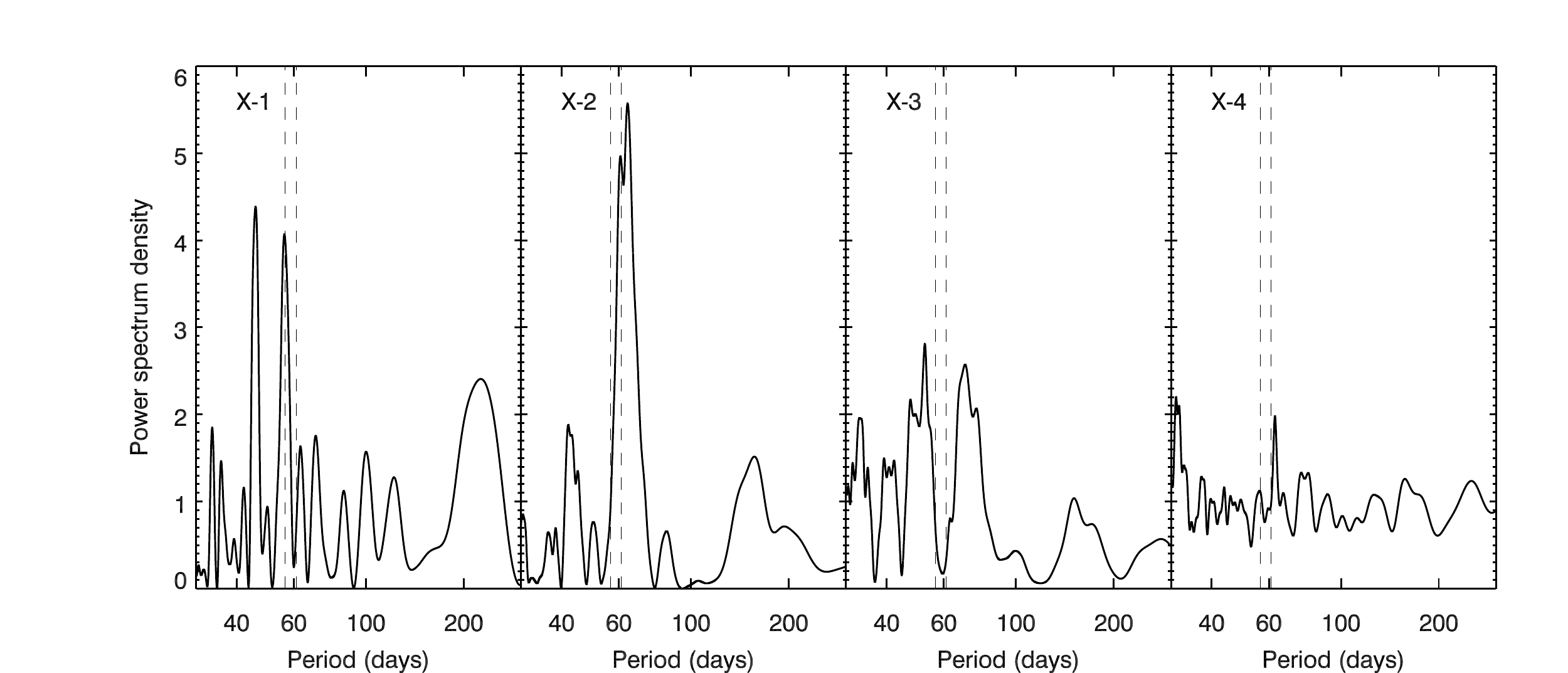}
\caption{Lomb-Scargle periodograms of the four brightest point sources in the center of M82 from the 2014-2016 \chandra\ data (black lines). The dashed lines mark the 61.0-day and 56.5-day periods detected from \swiftxrt\ monitoring.}
\label{fig_chandra_period}
\end{center}
\end{figure*}

As done for the \swiftxrt\ data, we conducted a LS periodogram analysis of the 2014-2016 \chandra\ data. We search over periods of 30--300 days, a narrower range of periods than in the \swiftxrt\ analysis since the cadence of the \chandra\ montitoring is longer and the duration is shorter. We plot these LS periodograms in Figure \ref{fig_chandra_period}. The most prominent peak in any of the periodograms is in that of X-2, at 63.8 days with a FWHM of 12.5 days.

To get an indication if the results obtained from the Lomb-Scargle periodogram can be independently confirmed, we fit the data with an alternative method, using a Gaussian process model. We use the implementation of these models called \textit{celerite}, in the \texttt{celerite} Python package\footnote{Here we use the same convention used by \cite{celerite}, where italic indicates the model and monospace indicates the software library. The library can be found at \href{http://celerite.readthedocs.io/en/stable/}{http://celerite.readthedocs.io/en/stable/}} \citep{celerite}.
In the figures in this Section, we show the results of the fitting of these models on the Chandra data of all four sources. 

The flux of X-1 and X-2 spans multiple orders of magnitude. In order to make the code run on the same kind of quantities it was designed for, we normalize the fluxes using a magnitude scale defined as $-2.5\,{\rm log}_{10}$(\fx$/10^{-10}$\ergcms), where \fx\ is the X-ray flux of the source in the 0.3-10 keV band. The exact choice of the reference flux does not influence the results.

The procedure we follow here is the following: We fit the data to a \textit{celerite} model with a single peaked component, initially set at 60 days but leaving it free to vary between 20 and 500 days. After an initial fit, we sample the posterior distribution of the parameters with a Monte Carlo Markov Chain in order to verify how stable the initial solution is. If the 60-day period is a statistical fluctuation, we expect the final MCMC step to show no evidence of it.

In \textit{celerite}, these quasi-periodic components are modeled as \textit{Single Harmonic Oscillator} terms and their power density spectrum is
\begin{equation}\label{eq:shot}
S(\omega) = \sqrt{\frac{2}{\pi}}\frac{S_0\,{\omega_0}^4}{(\omega^2 - {\omega_0}^2)^2 + {\omega_0}^2\,\omega^2/Q^2}
\end{equation}
where $\omega_0$ is the characteristic frequency of the oscillator, $Q$ is the quality factor, and $S_0$ is defined so that $S(\omega_0) = \sqrt{2/\pi} S_0 Q^2$. 

$Q$ measures how ``peaked'' is a component in the power spectrum: the quality factor is an estimate of the number of oscillations that are made before losing coherence. So, higher $Q$ values indicate longer-lasting oscillations. 

We follow the first example in \cite{celerite}, using our data instead of the simulated data. We adapt the Jupyter notebook used in the original paper to produce Figure 4 in that paper. We find the best-fit values of $Q$, $S_0$ and $\omega_0$ by minimizing a log likelihood function, as explained in the \texttt{celerite} manual and by \cite{celerite}. The initial priors for the parameters are: $3 < \log\,Q < 15$, with starting value $Q = 1000$ so that we are looking for actual long-lasting oscillations; period $25\,\mathrm{d} < P < 100\,\mathrm{d}$, starting value 30\,d (arbitrary, far from 60); $log(S_0)$ between -15 and 15, practically unconstrained. Optionally, we can add a stochastic noise component, with $Q=1/\sqrt{2}$ (fixed, to mimic a $1/f$-like noise component, as per \cite{celerite}), and the other parameters $-15 < \log\,S_N < 15$ and $-15 < \log\,\omega_N < 15$. The minimization is performed using the L-BFGS-B method, as implemented in \texttt{scipy}.

Finally, we run an MCMC sampler using the \texttt{emcee} library, in order to estimate some meaningful intervals for the parameters by sampling the posterior distribution, using the appropriate log-probability for the \textit{celerite} process. We create 5000 samples of the parameters after a burn-in phase where we throw away the first 500 steps. Due to the high number of results pegged at the lower limit of the period, we discard all results for periods $<$20\,d.

To estimate the noise levels, and to assess the significance of these results, we use two methods. First, we randomize the observing times and maintain the flux values and their errors, doing this 1000 times; then, we calculate the confidence limits from the 84\% and 97.5\% percentiles (equivalent to 68\% and 95\% confidence levels or 1 and 2-$\sigma$). Second we only scramble the flux measurements, which has the effect of retaining any imprint of the observing strategy. Again, we use the same percentiles above to evaluate the 1- and 2-$\sigma$ levels.

\begin{figure*}
\plottwo{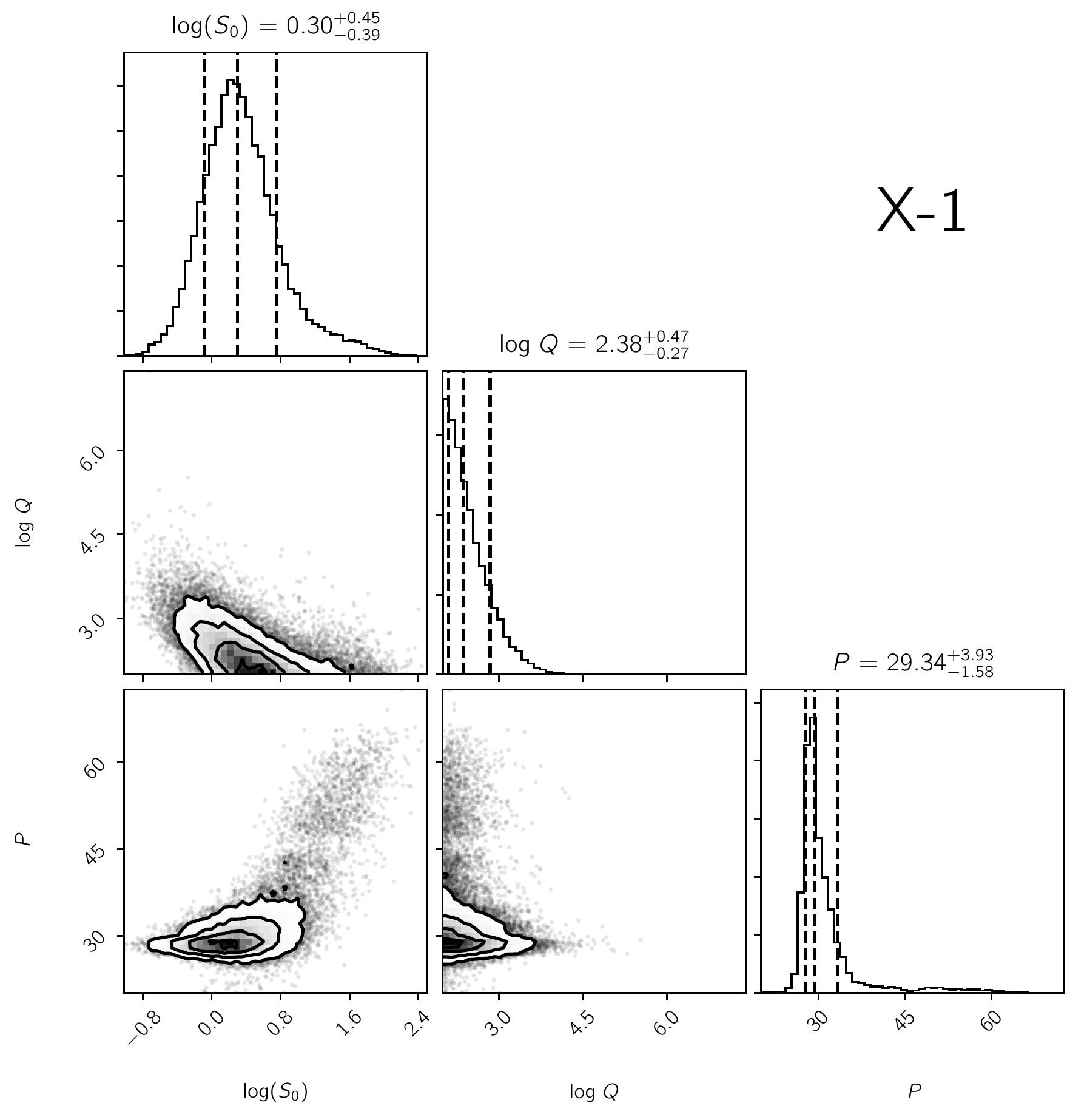}{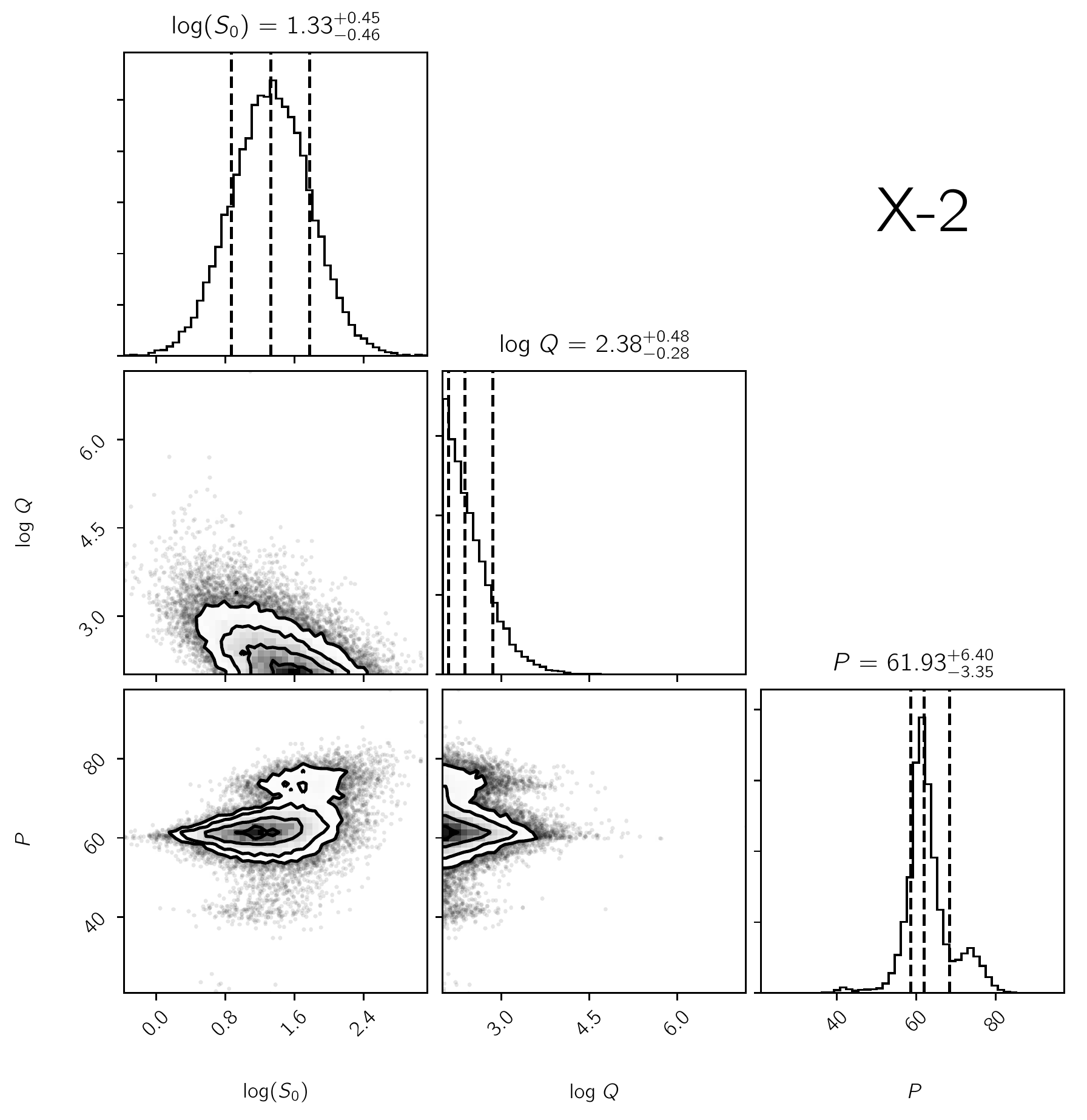}
\plottwo{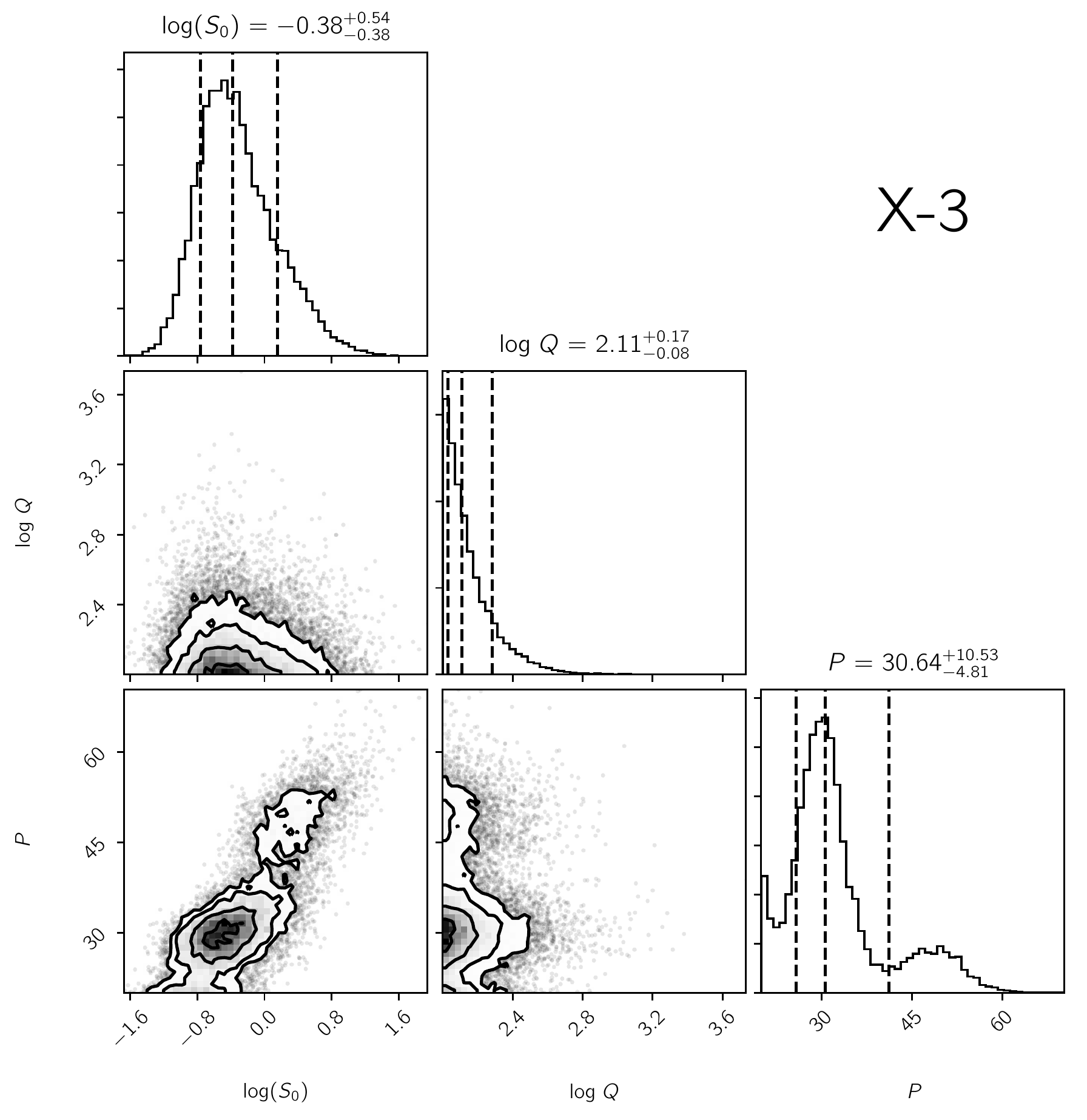}{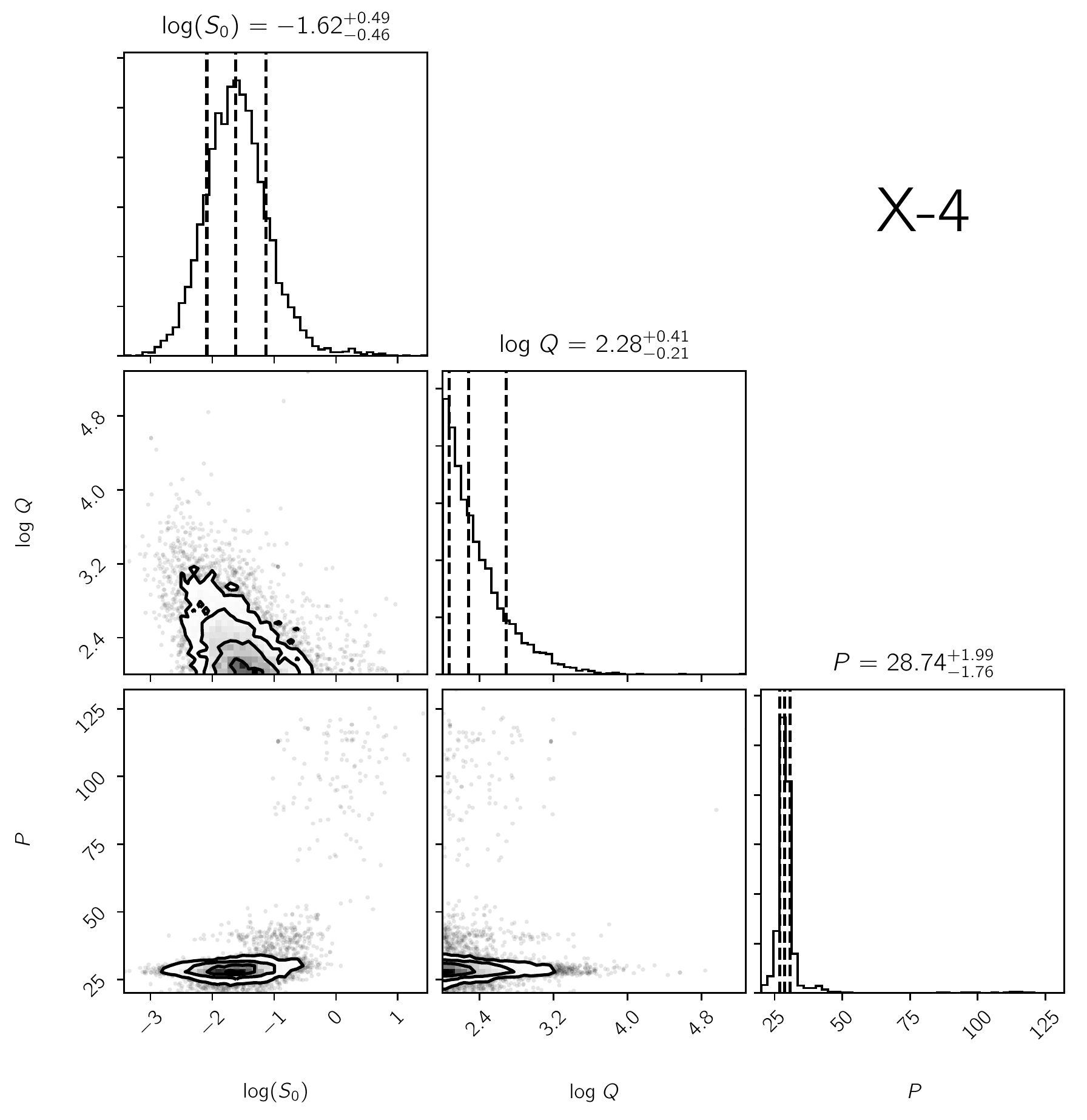}
\caption{Posterior probability distributions for the parameters of the \textit{celerite} model. 
The histograms along the diagonal show the marginalized posterior distributions for the single parameters.
Contours show the one, two and three-sigma confidence intervals for each pair of parameters.
The quoted uncertainties correspond to the 16 and 84\% percentiles of the marginal distributions.} 
\label{fig:corner}
\end{figure*}

\begin{figure*}
\plottwo{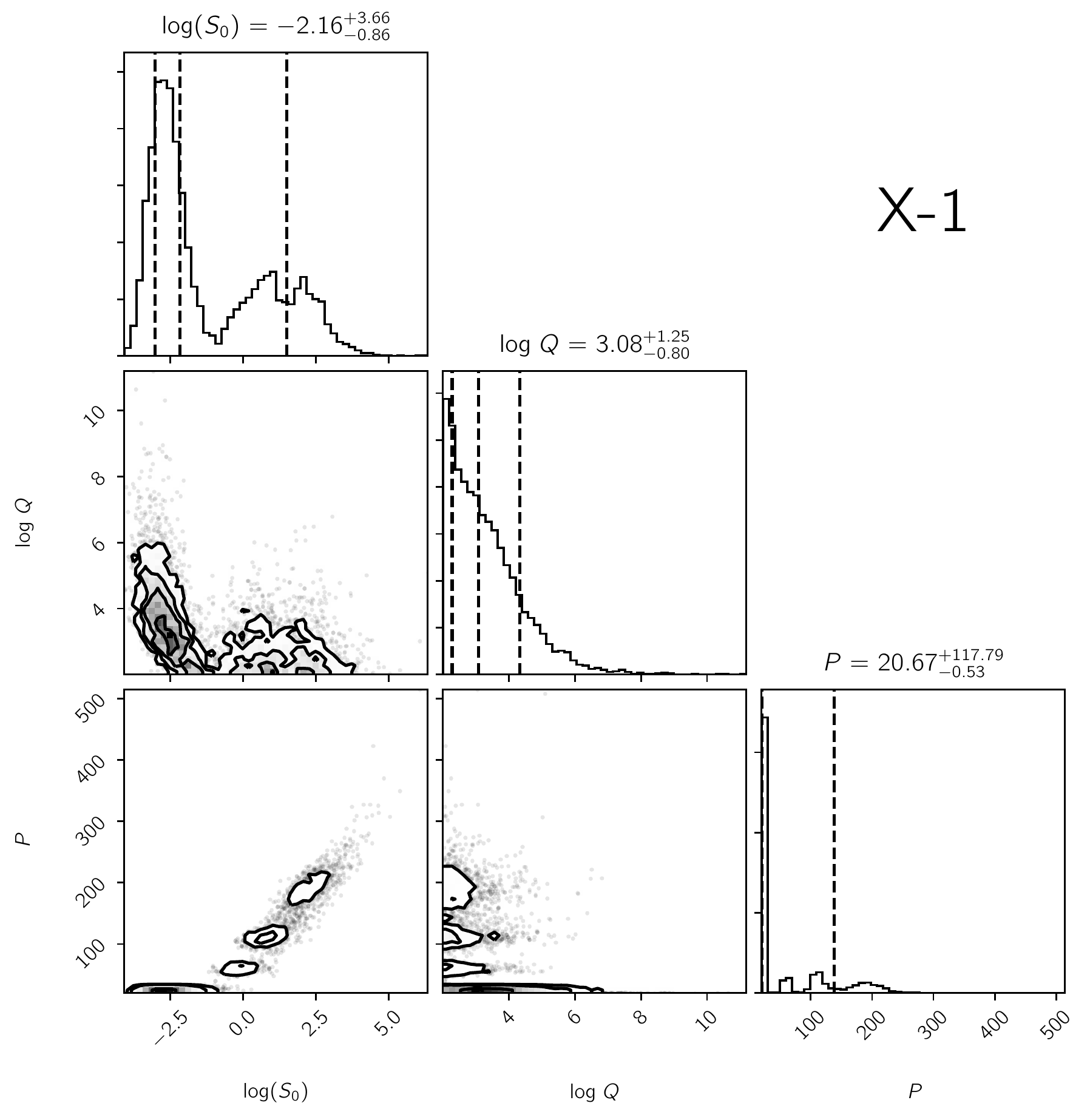}{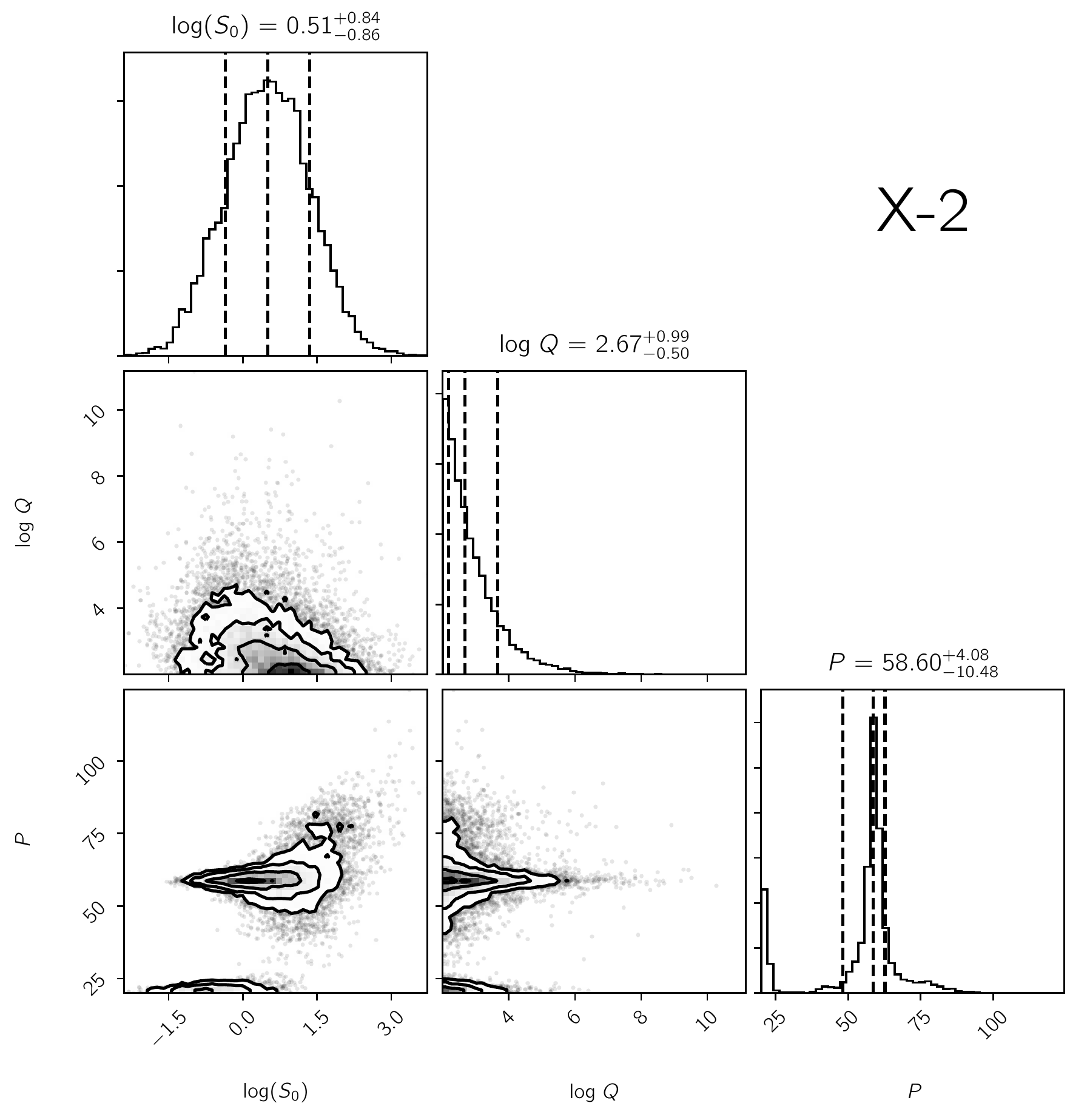}
\plottwo{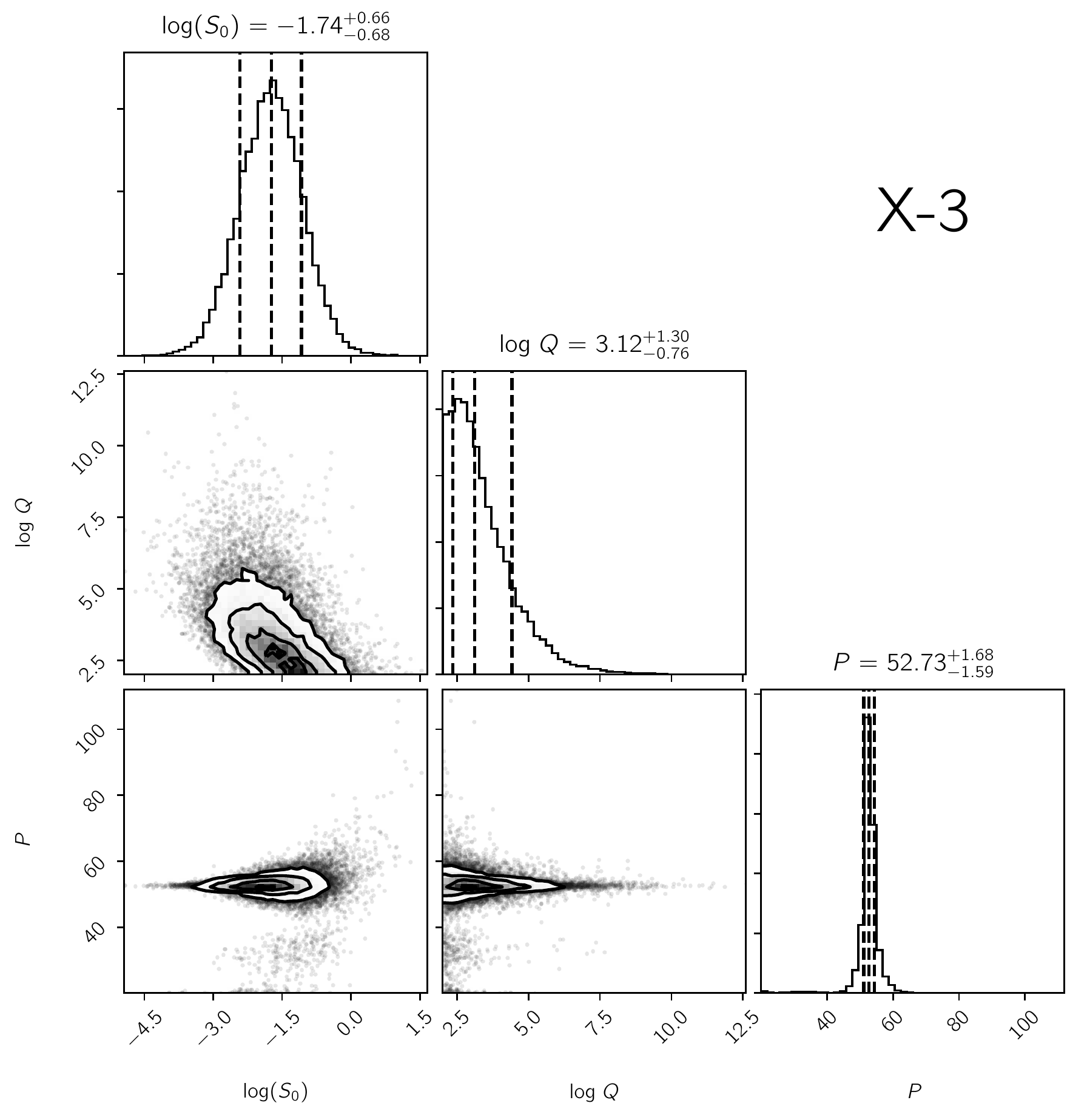}{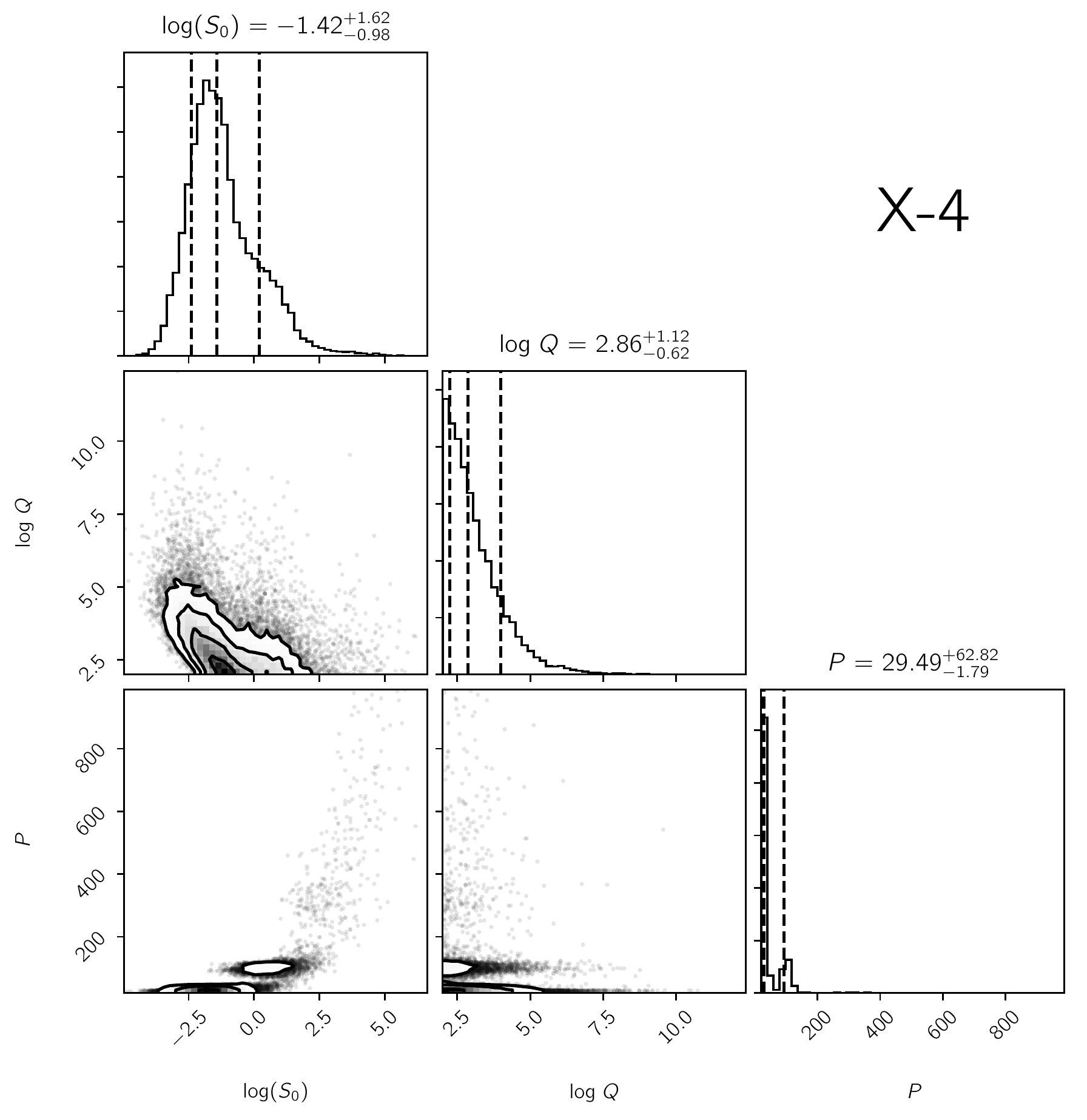}
\caption{Same as Fig.~\ref{fig:corner}, but limited to observations in the last 1000 days.}
\label{fig:cornerfilt}
\end{figure*}

\begin{figure*}
\includegraphics[width=\linewidth]{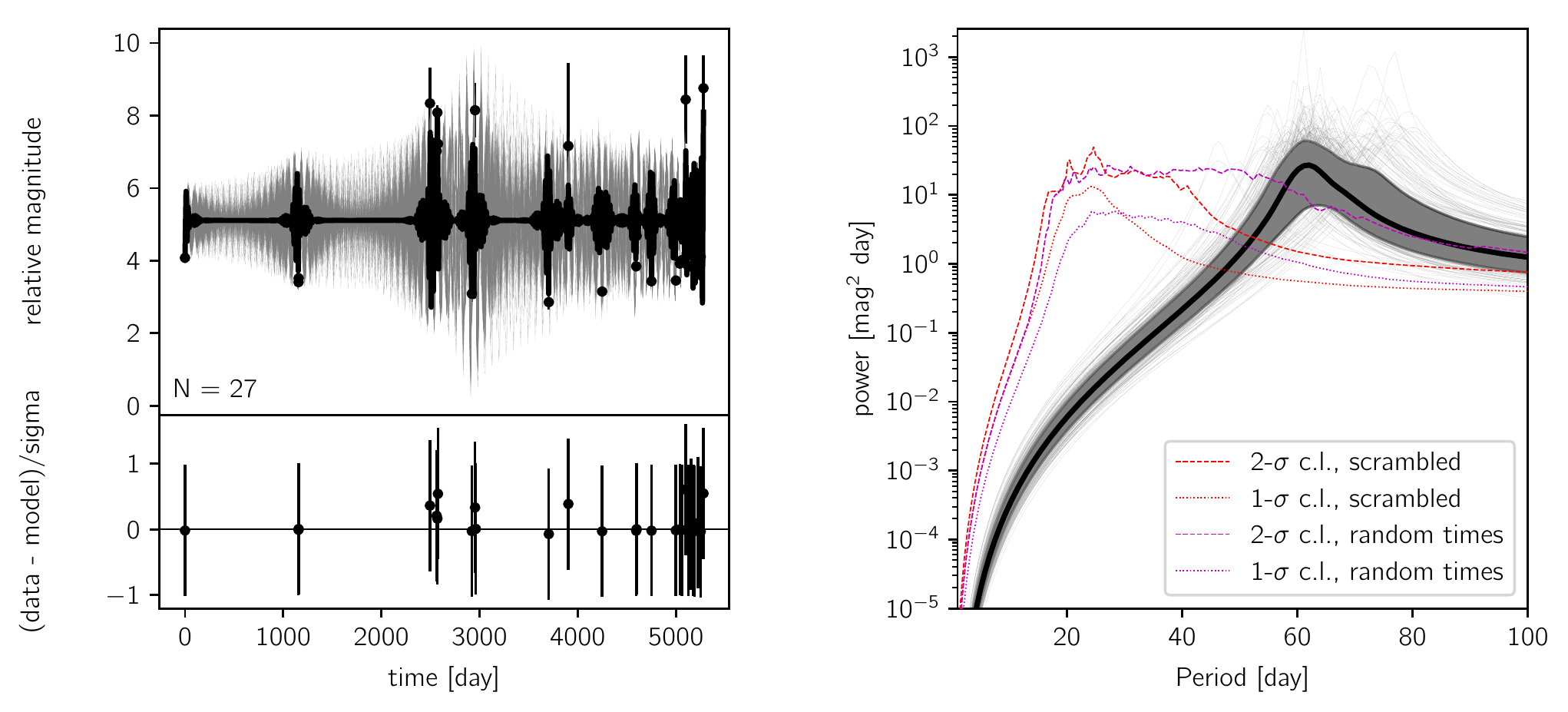}
\includegraphics[width=\linewidth]{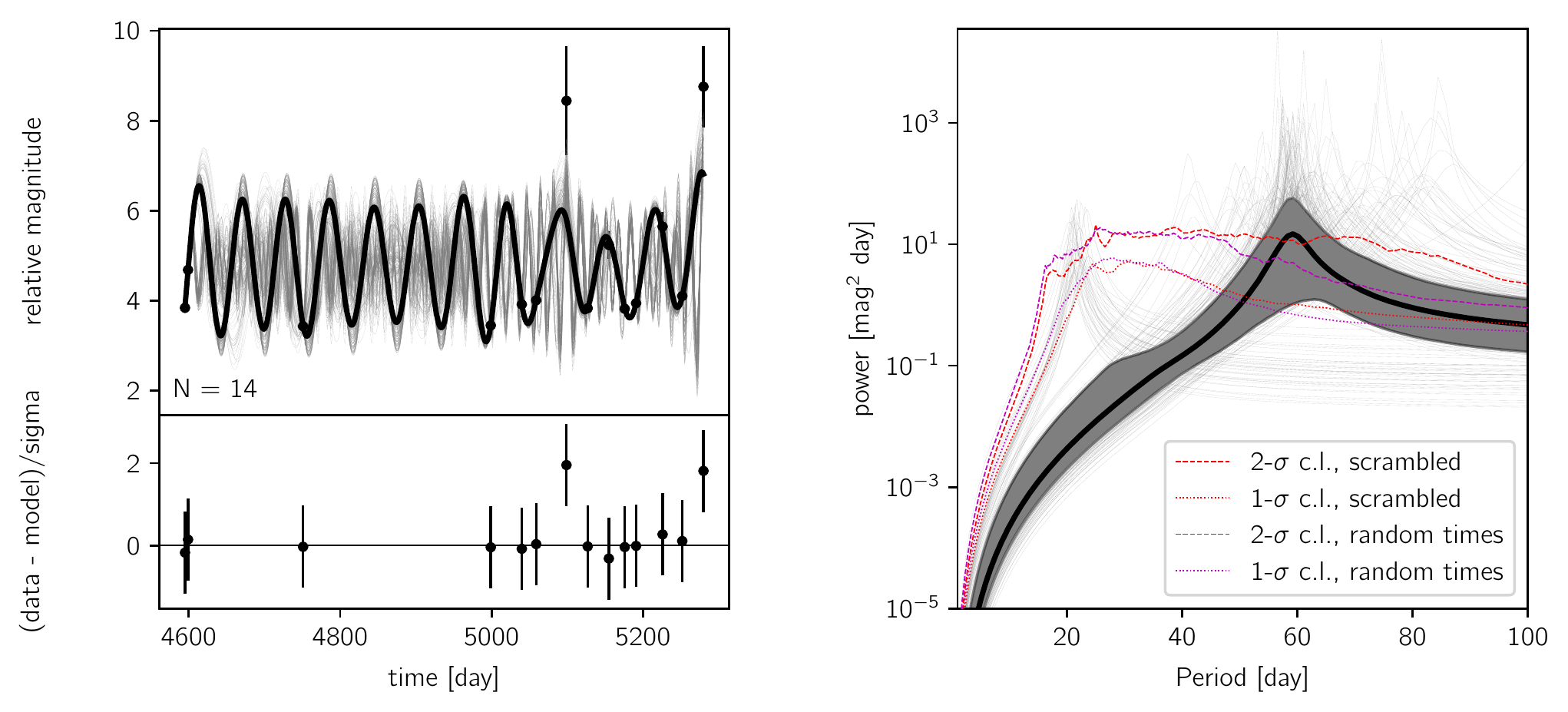}
\caption{Best-fit solution of the \textit{celerite} modeling of the light curve of X-2 done on Figures~\ref{fig:corner} and \ref{fig:cornerfilt},
(Top) on the full dataset and (Bottom) limited to the last 1000 days of observations.
On the left, we plot the light curve and the prediction from the Gaussian Process model.
On the right, the power spectrum.
The black line indicates the initial fit of the power spectrum and the light curve.
The grey, semi-transparent lines indicate 300 random realizations of the Gaussian Process, simulated in the MCMC.}
\label{fig:lc_x2}\end{figure*}

The results of the parameter distributions are reported in Figures~\ref{fig:corner} and ~\ref{fig:cornerfilt}, on the whole dataset and on the last 1000 days respectively.  The parameters are highly correlated to each other and the posterior distribution is not regular.  A period of $\sim60$\,d appears invariably in the first best-fit for X-2 and not for the other sources, and remains as a peaked feature in the posterior distribution as well. Figure \ref{fig:lc_x2} shows the best-fit lightcurve and best-fit power spectrum for X-2, again for the whole dataset and on the last 1000 days respectively. The noise levels from both methods as calculated above are also plotted which show that the $\sim60$-day signal from X-2 is in excess of the 2-$\sigma$ levels estimated from both. Approximately, we can say that the period found in X-2 is $\sim(60 \pm 10)\,d$ (1-sigma) in both the whole dataset and the last 1000 days of data.
The distribution of best periods is far from symmetric, with strong correlations with the other parameters.

\begin{figure*}
\includegraphics[width=\linewidth]{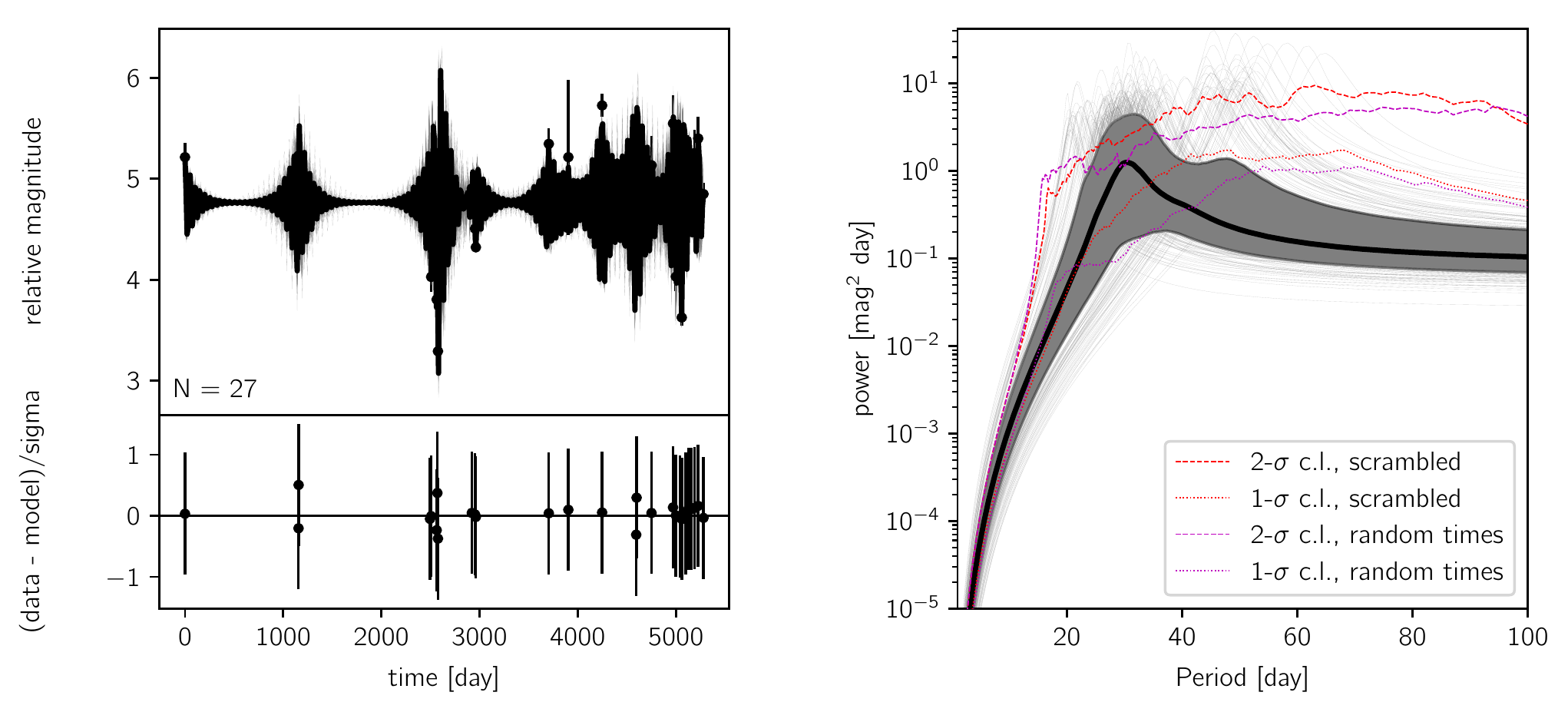}
\includegraphics[width=\linewidth]{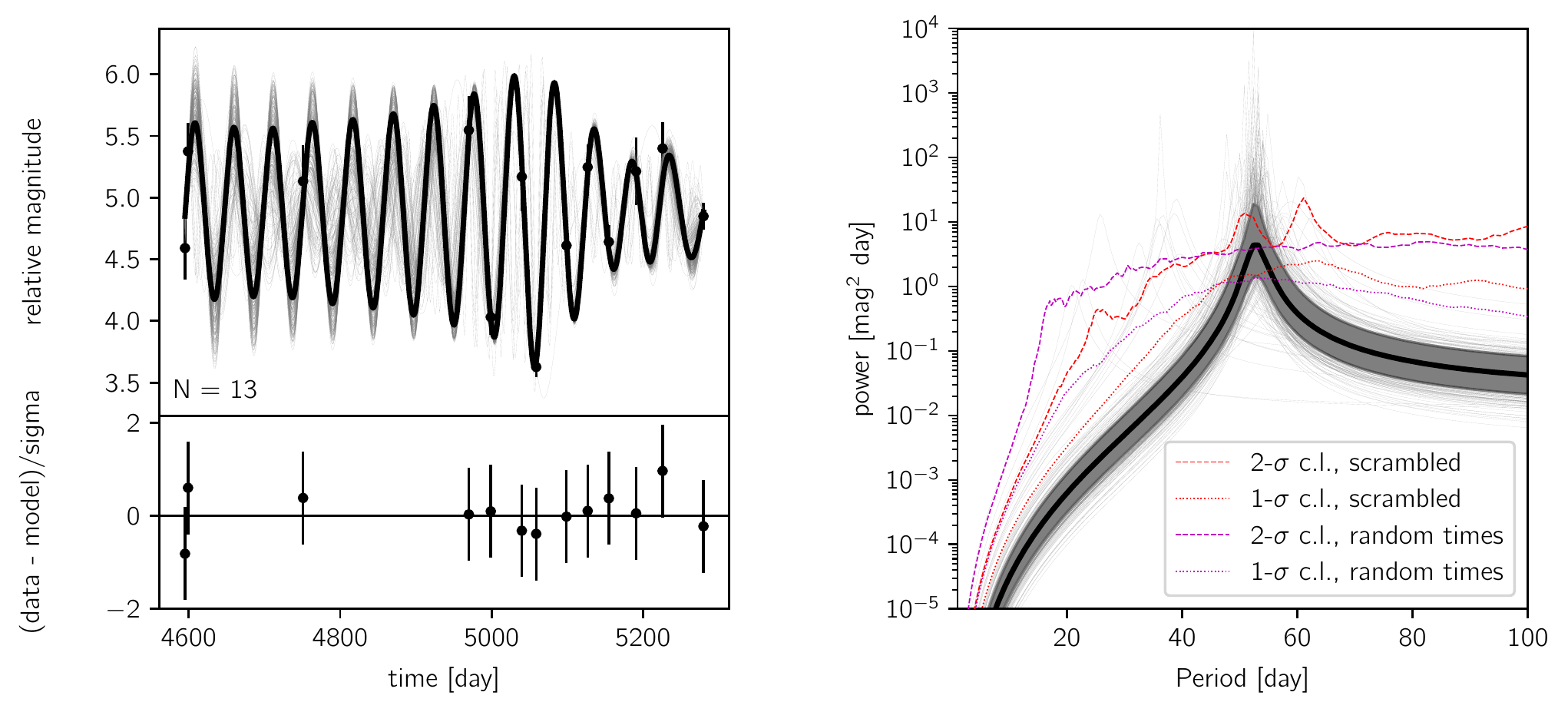}
\caption{Same as Fig.~\ref{fig:lc_x2}, but for X-3}
\label{fig:lc_x3}\end{figure*}

\begin{figure*}
\includegraphics[width=\linewidth]{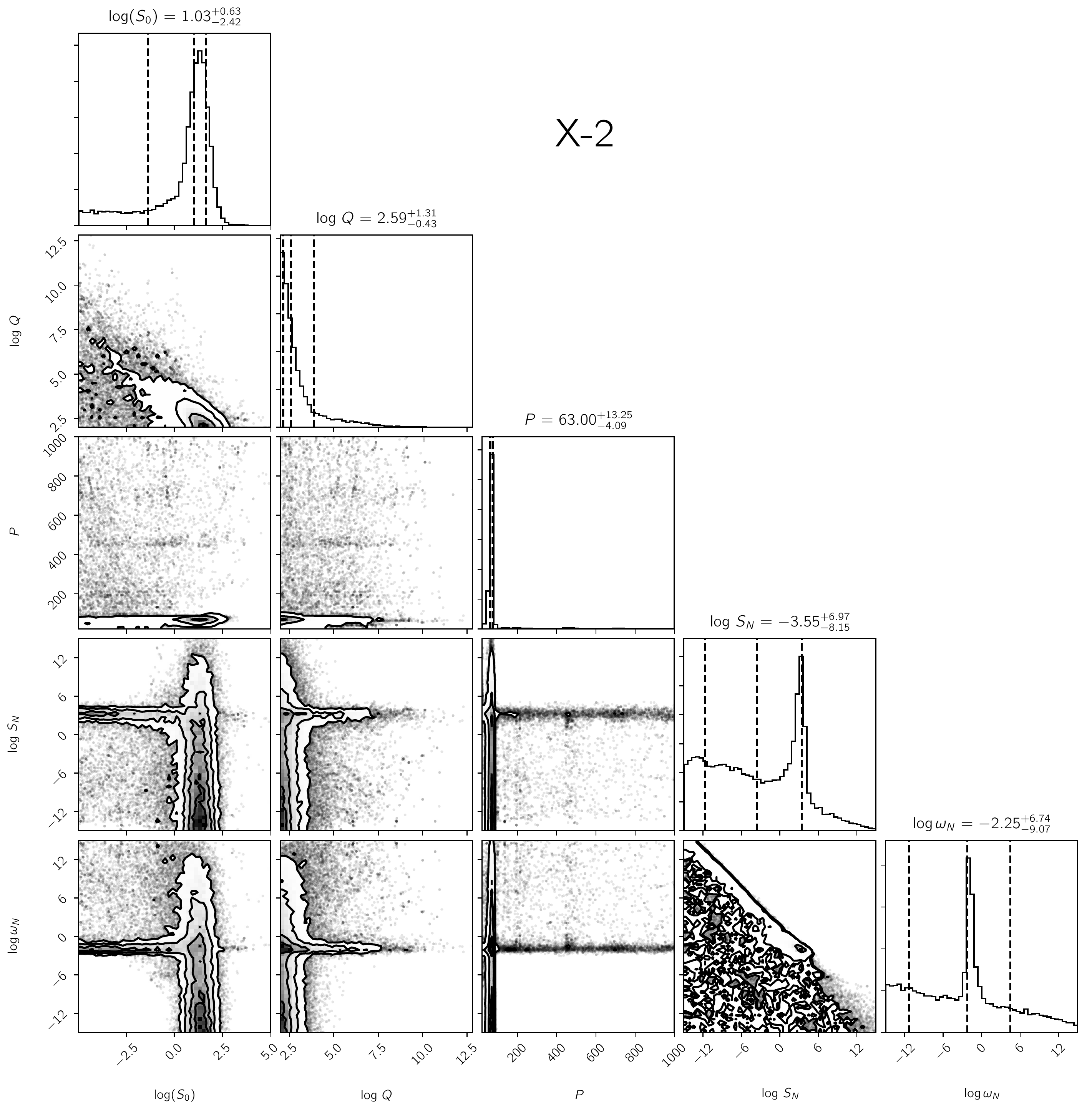}
\caption{Same as Fig.~\ref{fig:corner}, but adding a ``red noise'' component with $Q=1/\sqrt{2}$. 
There is a clear bimodality: if the red noise component is strong, the oscillatory component is damped, and vice versa.
This argues further against the combination of two sources of variability in the dataset.}
\label{fig:corner_rednoise}
\end{figure*}

Note that the scatter of residuals in Figures~\ref{fig:lc_x2} and \ref{fig:lc_x3} is lower than error bars, which is an indication that this model is probably an overkill, it is trying to overfit the data. Therefore, it comes to no surprise to see that adding a stochastic noise component creates a bimodal situation, where the peak at 60\,d appears where the amplitude of the stochastic noise component has a very low amplitude, and vice versa (Figure~\ref{fig:corner_rednoise}).

Nonetheless, as can be seen in all plots in this Section, X-2 is the only source where the peak around 60 days is fitted on the whole history of the source and on the time-filtered data, while in all other cases this narrow feature sets at different frequencies or does not appear at all in the whole dataset. This is confirmed by the posterior distribution sampling done in the MCMC.

Curiously enough, when filtering for the last 1000 days of data, a peak also appears in X-3 (Fig.~\ref{fig:lc_x3}) close to 53 days. 
Note however that the amplitude of the oscillation in the light curve of X-3 is lower than X-2.  This may or may not be an effect of the point spread function of X-2 partially overlapping with X-3 (See Figure \ref{fig_18064+18068_img}) and affecting the flux measurement. Since the periods are slightly different, and there is no obvious correlation between the fluxes measured in the two sources, we leave this to future analysis. X-1, X-3 and X-4 also exhibit a possible component at 30\,d, below 2-$\sigma$. If real, we are prone to attribute it to X-1, because the source is so bright that it might be influencing the flux measurements of these nearby and much weaker sources.

\section{Is X-2 responsible for the periodic modulation seen by \swiftxrt?}
\label{sec_period}

The $\sim$60-day periodic signal that we detect from X-2 in the 2016 \chandra\ data is similar to the 61.0-day and 56.5-day periodic signals detected by \swiftxrt\ from 2012--2104. However, during 2016, when X-1 was highly active, no periodic signal is detected from \swiftxrt, which is dominated by the flux from X-1, so a comparison of contemporaneously detected signals is not possible. We instead compare the average flux profiles of the signals detected by \swiftxrt\ that we calculated in Section \ref{sec_swift} to the observed \chandra\ fluxes of X-2. 

We allow for a change in normalization of the signal, which both accounts for an intrinsic change in normalization and for the integrated fluxes of the other X-ray sources included in the \swiftxrt\ data that cannot be spatially resolved and which vary over time. The change in normalization is accounted for by taking the average flux profiles and subtracting off a constant. We vary the constant between 0--2$\times10^{-11}$ \ergcms\ in steps of 2$\times10^{-13}$ \ergcms. We also allow for any possible phase changes that are known to occur. For each value of the constant subtracted off, we cycle through phase space in steps of one day. 

For each flux and phase value we test the goodness of fit of the derived profile of the observed periodic signal to the observed \chandra\ fluxes of X-2 using \chisq\ statistics. Since both quantities have uncertainties, we add these together, however they are dominated by the uncertainty in the flux profile. We do this for the flux profiles of the signals found in epoch 1 and epoch 2. 

\begin{figure}
\begin{center}
\includegraphics[width=90mm]{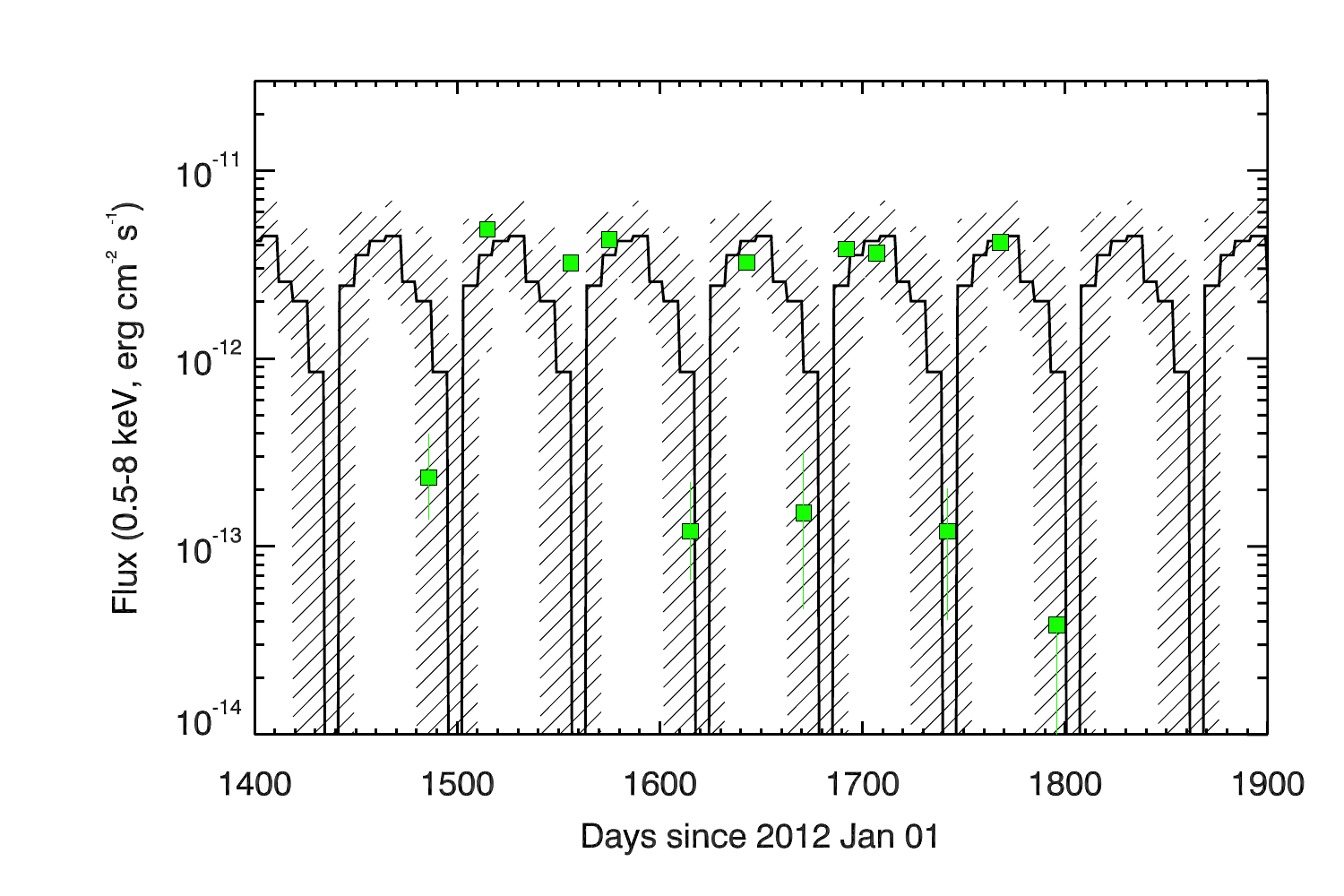}
\includegraphics[width=90mm]{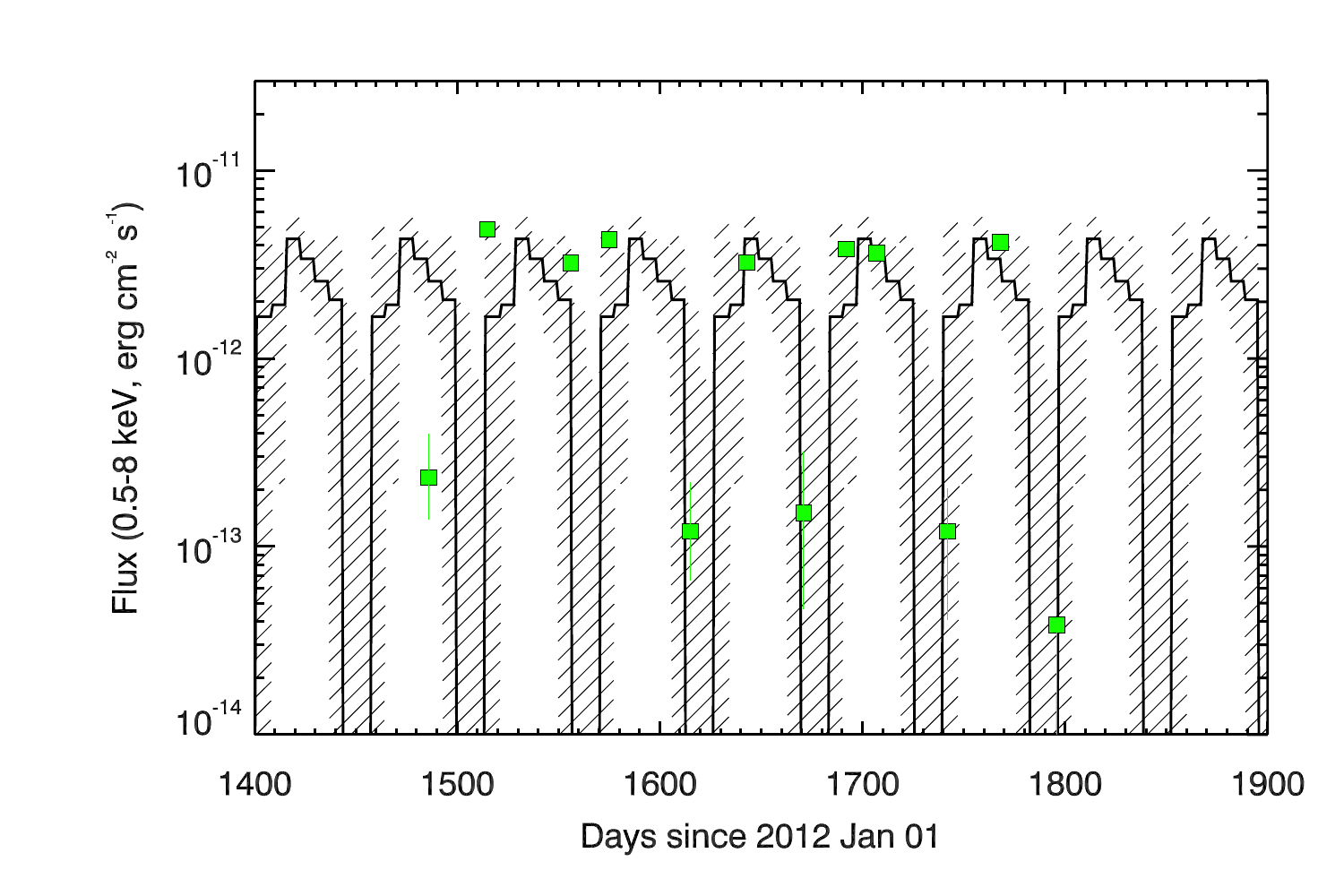}
\caption{2016 \chandra\ 0.5--8 keV lightcurve (green points) of X-2 compared to the mean flux profile of the 61.0-day signal (top, black solid line) and 56.5-day signal (bottom, black solid line) detected in \swiftxrt\ monitoring. For the 61.0-day signal a constant flux of 1.3$\times10^{-11}$ \ergcms\ has been subtracted and a phase shift of 31 days applied. For the 56.5-day a constant flux of 1.5$\times10^{-11}$ \ergcms\ has been subtracted and a phase shift of 52 days applied. The hatched regions in both panels represent the 1$\sigma$ spread in the \swiftxrt\ flux profiles.}
\label{fig_X2_period}
\end{center}
\end{figure}

We find that for epoch 1 the minimum \chisq\ (\chisq=2.4 for 10 degrees of freedom) is found when a constant flux of 1.3$\times10^{-11}$ \ergcms\ is subtracted and the phase has been shifted by 31 days (a phase shift of $\sim0.5$). For epoch 2, the minimum \chisq\ (\chisq=6.5 for 10 degrees of freedom) is found when a constant flux of 1.5$\times10^{-11}$ \ergcms\ is subtracted and the phase is shifted by +52 days or -4 days. We show these mean profiles with the best-fitting flux subtracted and phase shift applied in Figure \ref{fig_X2_period}. From the \chisq\ values and this figure, we find that the flux variability profile of X-2 we observe in 2016 is fully consistent with the periodic signal seen by \swiftxrt, with a change in phase. 

We also test if the observed \chandra\ fluxes of X-1, X-3 or X-4 could also account for the observed modulation. Carrying out the same profile fitting that we conducted for X-2, we find that the mean profile also compares well with the \chandra\ data on X-3. However, since no periodicity around 60-days is detected in the LS periodogram of these data (only a peak around 20--30 days is seen, Figure \ref{fig_chandra_period}), we find it unlikely that X-3 is the source of the signal. The periodogram of X-1 also presents a peak at $\sim60$ days that is just above the noise curve, however, since no similar peak is detected in the \swiftxrt\ data of epoch 3, where X-1 is dominating the flux, we also consider it unlikely that X-1 is the source of the signal. This is in agreement with \cite{qiu15} who find no periodicity originating from region around X-1. While they could not determine exactly which source the signal did originate from due to the lack in spatial resolution of \swiftxrt, our new \chandra\ data conclusively point to X-2 as the origin.

M82 had been observed by \chandra\ on many previous occasions since its launch in 1999 prior to our new monitoring campaign. These data were described and analyzed in \cite{brightman16}. They consisted of a heterogeneous mix of exposure times, off-axis angles, and instrumental set up, and were taken at irregular intervals, and thus were not conducive to a search for periodicities. Our new data have the advantage over the archival data in that it consists of a homogeneous data set taken regularly at intervals with the same instrumental set up.  While we do not detect a peak around 60 days in a LS test of the archival data, the more sophisticated auto-regression analysis shows that the periodicity is stable over the 17-years of observations (Fig \ref{fig:lc_x2}).

\begin{figure}
\begin{center}
\includegraphics[width=90mm]{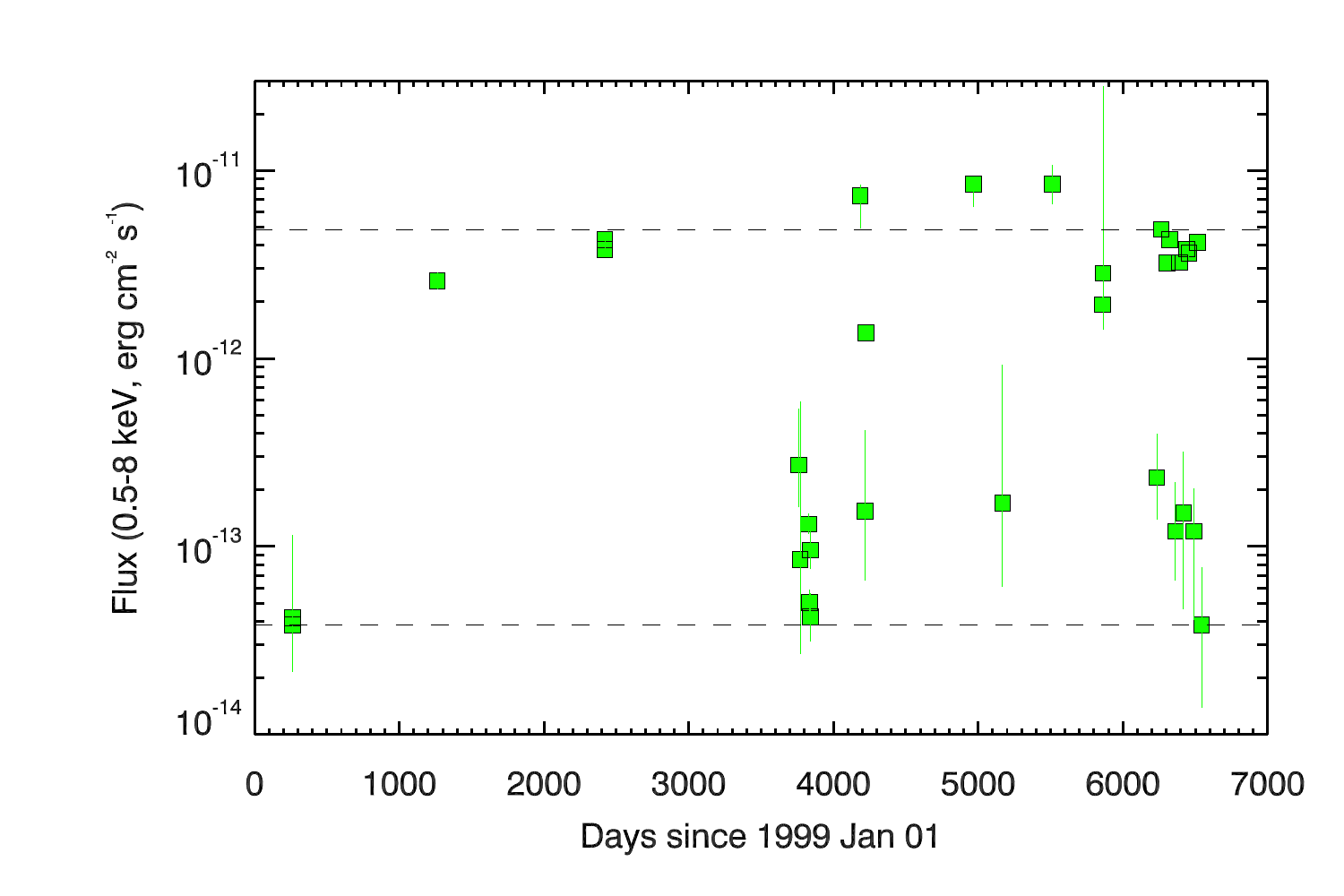}
\includegraphics[width=90mm]{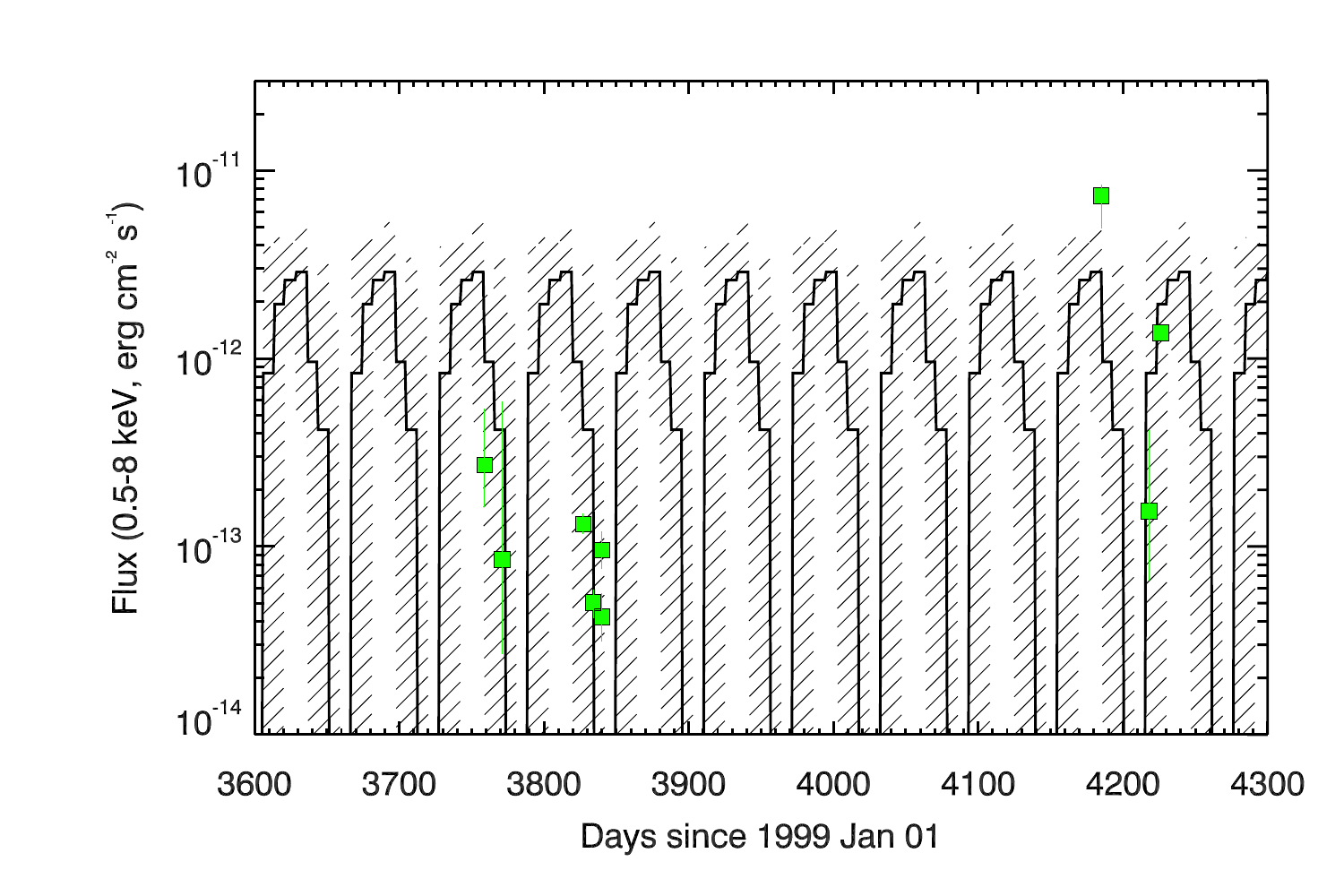}
\caption{Top: 1999--2016 \chandra\ 0.5--8 keV lightcurve (green points) of X-2. The dashed lines represent the maximum and minimum fluxes that we observed during the 2016 monitoring campaign. This shows that with the exception of three observations, the flux ranges observed in 2016 were representative of the long-term range. Bottom: Zoom in to the 9 observations taken in the period 2009-04--2010-08 with the best-fit \swiftxrt\ flux profile overlaid.}
\label{fig_X2_ltcrv}
\end{center}
\end{figure}

We show a long-term light curve of our data combined with the archival data in Figure \ref{fig_X2_ltcrv}. This shows that the flux range exhibited by the archival data were very similar to our new data. Dashed lines mark the maximum and minimum flux levels observed in our new data, and the old data almost all fall inside this range, with the exception of three data points that appear to lie above the maximum flux. These data were however subjected to the effects of pileup, and hence the flux may have been overestimated. See \cite{brightman16} for more details. Alternatively however, it was during one of these epochs that the pulsations were first detected (at $\sim5500$ days), so the increased brightness at this time could have come from the pulsed component which has not been detected since (Bachetti et al. in preparation).

We also run a fit to the profile of the \swiftxrt\ profile on the long-term light curve in the same way as described above. Since the data are spread over 16 years, and phase changes are likely to have occurred, we limit this analysis to a period over which 9 observations were take over the period of $\sim$1.3 years, from 2009-04--2010-08. We show the fit in Figure \ref{fig_X2_ltcrv}. This shows that likewise for our new data, the long-term \chandra\ fluxes of M82 X-2 can be fully described by the periodically modulated \swiftxrt\ flux profile.

\section{Implications for the ULX pulsar}
\label{sec_implications}

We have unambiguously identified the ULX pulsar, M82 X-2, as the source of the $\sim$60-day periodicity originating from M82, first discovered by {\it RXTE} \citep{kaaret06} and confirmed by \swiftxrt\ \citep{qiu15} from a systematic monitoring by \chandra. From the \chandra\ monitoring data, which extended over a period of $\sim1$ year, we find the period to be 63.8$\pm0.6$ days. Since the orbital period of the system has been determined to be 2.5 days \citep{bachetti14}, the 63.8-day period must be super-orbital in origin. The modulations are roughly sinusoidal with a peak-to-trough amplitude of $\sim5\times10^{-12}$ \ergcms. The \chandra\ minimum and maximum fluxes corresponds to a factor $\sim$100 variation in flux. 

The period of the super-orbital flux modulation from M82 X-2 bears a striking similarity to the periods of the flux modulations observed from the other ULX pulsars, NGC~7793~P13, where the period is 65.1 days \citep{hu17} and NGC~5907~ULX1, where the period is 78.1 days \citep{walton16b}. However, while in these systems the flux is only observed to vary by a factor of $\sim2-3$, the flux from M82 X-2 varies by a factor that is far larger. Furthermore, both NGC~7793~P13 and NGC~5907~ULX1 exhibit `off' states where their fluxes are significantly lower than expected given an extrapolation of the periodic signal \citep{motch14,walton15b,walton16b,fuerst16}. We find no evidence for any additional off states from M82 X-2 lower than those caused by the super-orbital modulations.

Super-orbital periods are known in several other well-studied neutron star binary systems such as Her X-1 \citep{tananbaum72}, LMC X-4 \citep{lang81}, and SMC X-1 \citep{gruber84}. In most cases, a precessing warped accretion disk model is the favored one, where the variations in intensity are caused by the warp in the accretion disk periodically occulting the neutron star opposed to by projection effects as discussed above. The super-orbital flux variations from these neutron star binaries are quite similar to those observed in M82 X-2. For example, for LXC X-4 the variations have a period of 30 days, are roughly sinusoidal in shape and vary by a factor of up to 60, dropping to zero during the minima.

One test of the warped accretion disk model for M82 X-2 is the expectation of strong spectral variations with flux. In \cite{brightman16} we explored the dependence of the absorption and spectral slope on the observed luminosity, however we did not find any strong evidence for a dependence of either on the luminosity. This was made challenging, however, since at its lowest fluxes, the brightness of X-2 is at a similar level as the local background, produced by bright diffuse emission, making it difficult to get good spectral constraints. 

\begin{figure}
\begin{center}
\includegraphics[width=90mm]{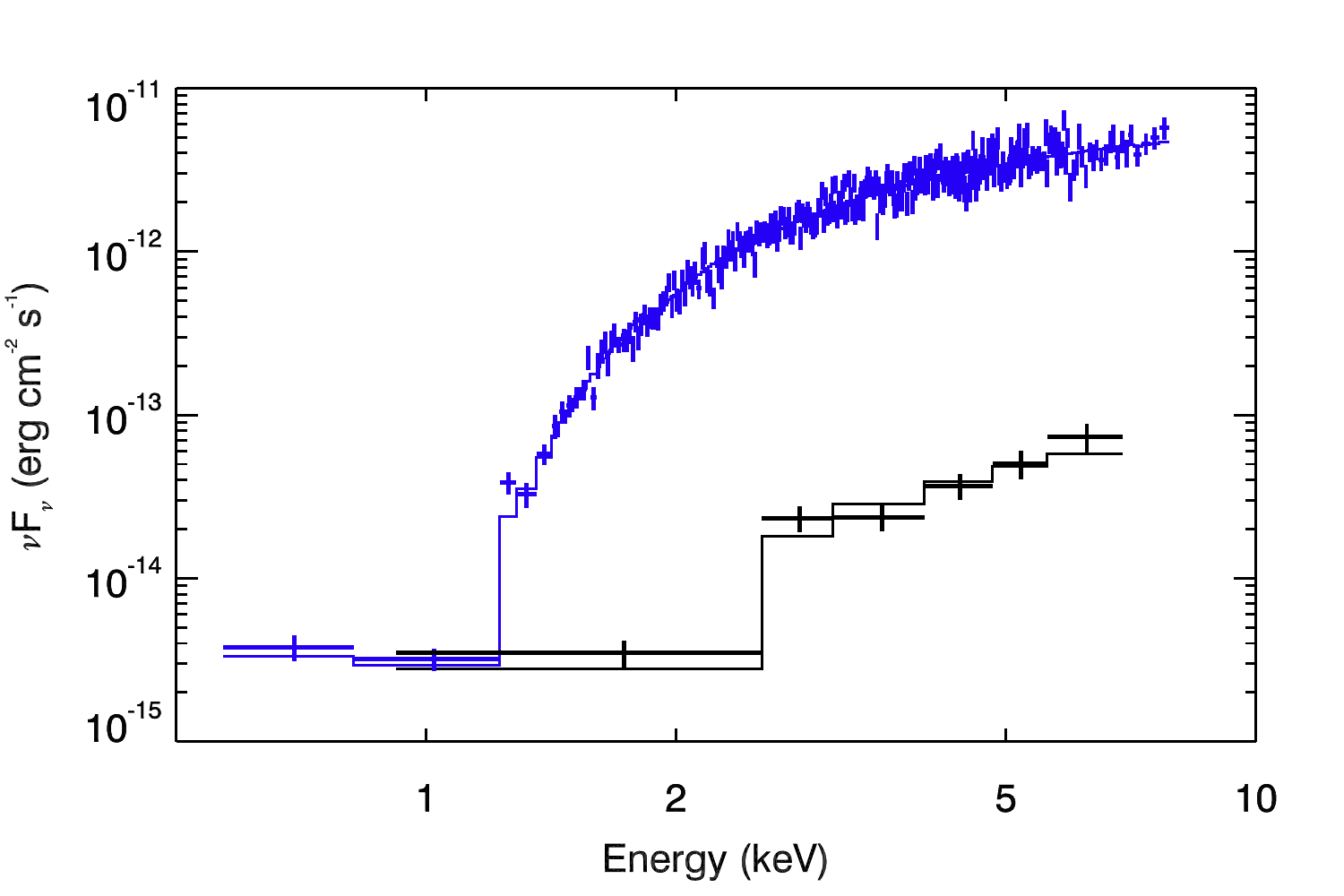}
\caption{Comparison of the spectral shape of X-2 at its maximum flux (blue) and minimum flux (black). The spectral shape is very similar as are the derived parameters}
\label{fig_X2_spectra}
\end{center}
\end{figure}

Fortuitously a 120-ks on-axis observation of M82 was made in 2009 (obsID 10543) where X-2 was at its minimum flux. Since the on-axis PSF is less than 1\arcsec, this observation gives us the best opportunity to study the spectral shape of X-2 at this low flux due to the low background inclusion. We found that the spectral shape was remarkably similar to that of the high state (e.g. obsID 5644), with $\Gamma_{\rm high}=1.42^{+0.05}_{-0.05}$ and $\Gamma_{\rm low}=1.20^{+0.57}_{-0.64}$. We show the spectra in Figure \ref{fig_X2_spectra}. We therefore find it highly unlikely that the flux variations are caused by occultations.

\cite{middleton18} invoked the precession of an accretion disk induced by the Lense-Thirring precession of a large-scale height accretion flow, to explain the long-term periodic flux modulations seen in ULX pulsars. Here projection effects rather than occultation effects are invoked to explain them. Furthermore, for NGC~5907~ULX-1, \cite{fuerst17} found evidence for spectral variations as a function of super-orbital phase which were attributed to a precessing accretion disk where the angle between our line of sight and the axis of the accretion disk changes. However, a precessing disk would require very large changes in inclination angle (i.e. from face-on to edge-on) to cause the almost 2-orders-of-magnitude variations seen from M82 X-2.

While the precession of the disk itself may not be the cause of the variations, precessing of a beamed component might. For example, \cite{dauser17} conducted Monte-Carlo simulations of large scale-height disk that is expected to form from super-critical accretion, and found that geometrical beaming by a small conical opening and precession can describe the observed flux variations NGC~5907~ULX-1. This model can produce a flux boost by a factor of 100. However, this scenario needs very fine tuning as it requires half-opening angles of the flow of 5\degree\ or less and a line-of-sight viewing angle that is very similar to the opening angle such that the flux drops dramatically when the beamed component precesses out of the line of sight. Furthermore, accretion flow models at high accretion rates have found that the disk always remains slim ($H/R<1$), precluding such thick accretion disks \citep[$H/R>1$,][]{beloborodov98,lipunova99,poutanen07,lasota16}.
 
Some previous works have interpreted the flux profile of X-2 to be bimodal, with `high' and `low' states possibly related to the propellor effect \citep[e.g.][]{tsygankov16}. However the propellor effect is related to changes in the accretion rate, and since the flux modulation is periodic, it is hard to understand how changes in the accretion rate can be periodic unless they are related to the binary orbit, which is both highly circular and at 2.5 days is far shorter than the 63.8-day periodic modulation.

Conversely, \cite{dallosso16} interpret the observed fluxes from X-2 as being a continuous distribution though their magnetically threaded disk model that describes the disk transitioning from radiation-pressure dominated at high fluxes, to gas-pressure dominated at low fluxes, however, again the emission according to their model depends on accretion rate, which is hard to reconcile with the long-timescale periodicity. Furthermore, spectral changes would be expected as a function of luminosity.

\section{Summary and Conclusions}
\label{sec_conclusions}

We have presented results on a systematic monitoring campaign of M82 with \chandra, with the goal of unambiguously determining the source of the $\sim60$-day periodic signal detected by {\it RXTE} and \swiftxrt. From a simple Lomb-Scargle periodogram analysis  and a more sophisticated auto-regressive moving average analysis of the flux variations from the four bright point sources in the center of M82,  we find that only the ULX pulsar, X-2, exhibits a signal at or around 60 days, from both the monitoring data and the longer-term archival data. We therefore confirm that this is the source of the well known periodicity from M82. We constructed a mean flux profile of the \swiftxrt\ signal and compare to the \chandra\ fluxes from X-2 and find that the observed \chandra\ fluxes from X-2 can fully explain the \swiftxrt\ flux profile once the other sources included in the \swiftxrt\ PSF and a change of phase are accounted for. The flux modulations with a $\sim60$-day period are at far larger time scales than the binary orbit and must therefore be super orbital in origin. The flux varies by a factor of $\sim$100 from minimum to maximum, with no evidence for spectral variations. We discuss several possible mechanisms to produce these observations, however none of these are capable of fully explaining them all.

\section{Acknowledgements}

We would like to thank the anonymous referees and the statistics referee for their critical reviews and input on this paper which improved it greatly. We would also like to thank Matthew Graham for valuable insight and discussion. M. Bachetti  acknowledges support from the Fulbright Scholar Program and A. Zezas acknowledges funding from the European Research Council under the European Union's Seventh Framework Programme (FP/2007-2013)/ERC Grant Agreement no. 617001. This project has received funding from the European Union's Horizon 2020 research and innovation program under the Marie Sklodowska-Curie RISE action, grant agreement no. 691164 (ASTROSTAT). The scientific results reported in this article are based on observations made by the \chandra\ X-ray Observatory. Support for this work was provided by the National Aeronautics and Space Administration through \chandra\ Award Number GO6-17080X issued by the \chandra\ X-ray Center, which is operated by the Smithsonian Astrophysical Observatory for and on behalf of the National Aeronautics Space Administration under contract NAS8-03060. This research has made use of software provided by the Chandra X-ray Center (CXC) in the application package {\sc ciao}. We also acknowledge the use of public data from the {\it Swift} data archive.

{\it Facilities:} \facility{\chandra\ (ACIS), {\it Swift} (XRT)}

\end{document}